\documentclass[letterpaper, 10 pt, conference]{ieeeconf}  % Comment this line out if you need a4paper

\IEEEoverridecommandlockouts                              % This command is only needed if 
\usepackage{graphics} % for pdf, bitmapped graphics files
\usepackage{epsfig} % for postscript graphics files
\usepackage{mathptmx} % assumes new font selection scheme installed
\usepackage{times} % assumes new font selection scheme installed
\usepackage{amsmath} % assumes amsmath package installed
\usepackage{amssymb}  % assumes amsmath package installed
\usepackage{subcaption}
\usepackage{float}
\usepackage{multirow}
\title{\LARGE \bf
A Contrastive Federated Semi-Supervised Learning Intrusion Detection Framework for Internet of Robotic Things
}

\author{Yifan Zeng$^{1}$ 
\thanks{This work was not supported by any organization}% <-this % stops a space \\School of Computer Science and Engineering\\ Guangzhou 510006, People’s Republic of China\\
\thanks{$^{1}$Yifan Zeng is with School of Computer Science and Engineering, Sun Yat-sen University, 510006 Guangzhou, People’s Republic of China
        {\tt\small zengyf53@mail2.sysu.edu.cn}}%
}%

\begin{document}

\maketitle
\thispagestyle{empty}
\pagestyle{empty}

%%%%%%%%%%%%%%%%%%%%%%%%%%%%%%%%%%%%%%%%%%%%%%%%%%%%%%%%%%%%%%%%%%%%%%%%%%%%%%%%
\begin{abstract}

In intelligent industry, autonomous driving and other environments, the Internet of Things (IoT) highly integrated with robotic to form the Internet of Robotic Things (IoRT). However, network intrusion to IoRT can lead to data leakage, service interruption in IoRT and even physical damage by controlling robots or vehicles. This paper proposes a Contrastive Federated Semi-Supervised Learning Network Intrusion Detection framework (CFedSSL-NID) for IoRT intrusion detection and defense, to address the practical scenario of IoRT where robots don't possess labeled data locally and the requirement for data privacy preserving. CFedSSL-NID integrates randomly weak and strong augmentation, latent contrastive learning, and EMA update to integrate supervised signals, thereby enhancing performance and robustness on robots' local unlabeled data. Extensive experiments demonstrate that CFedSSL-NID outperforms existing federated semi-supervised and fully supervised methods on benchmark dataset and has lower resource requirements.

\end{abstract}

%%%%%%%%%%%%%%%%%%%%%%%%%%%%%%%%%%%%%%%%%%%%%%%%%%%%%%%%%%%%%%%%%%%%%%%%%%%%%%%%
\section{INTRODUCTION}

\textbf{Background.} Today, automation technology and robotic systems are widely deployed in industrial and commercial sectors. And robotics technology can deeply integrate with the IoT, forming IoRT, where robots are interconnected through network \cite{IoRT1}. This creates a new intelligent network infrastructure made up of robots and other automation devices as edge nodes. For instance, in the intelligence industry, IoRT enables various industries to employ multiple networked robots and other automation devices working collaboratively to handle tasks, achieving industrial automation and boosting efficiency \cite{Robotresource}. As IoRT devices, robots and other automation devices employ sensors, actuators, and wireless communication modules to understand environments, respond accordingly, and connect to network \cite{IoRTsensor}. IoRT provides remote access, enabling managers to remotely monitor and control robots and other automation devices.

However, network intrusion in IoRT poses risks, including data leakage, service disruption, and even illegal control of robots and other automation devices leading to serious physical harm. Sensitive information like  industrial secrets can be leaked, causing financial losses, and reputation damage. Disrupted services impact critical infrastructure and industrial processes, leading to accidents, downtime, and environmental hazards. Remote control of vehicles, robots and other automation devices by intruders threatens public safety and result in catastrophic outcomes. 

\textbf{Motivation.} Currently, Deep Learning-based Network Intrusion Detection System (DLNIDS) serves as an effective and automated defense measure \cite{cui2023novel}. When IoRT network intrusion detection system (NIDS) equipped in robot's network module detects IoRT traffic in attack category, it can trigger an alarm to remind administrators or automatically take measures against abnormal traffic connections based on programs. By precisely classifying attack traffic into more detailed and specific attack categories, the IoRT NIDS can take more targeted measures to defend against intrusions.

However, IoRT devices are predominantly robots or other automation devices, which have limited computing and storage resources \cite{Robotresource}. IoRT NIDS equipped in robotic and automation devices should be lightweight and real-time. Furthermore, obtaining labeled intrusion traffic data is highly time-consuming and expensive, making it challenging to label data at robot clients. Consequently, training DLNIDS locally on robots is difficult. Additionally, IoRT operates in such as industry, commerce, or military, where communication-generated traffic data may contain sensitive information. Uploading such data to a cloud server for training DLNIDS could potentially lead to data breaches.
\vspace{-10pt}
   \begin{figure}[H]
      \centering
      \includegraphics[width=8.0cm]{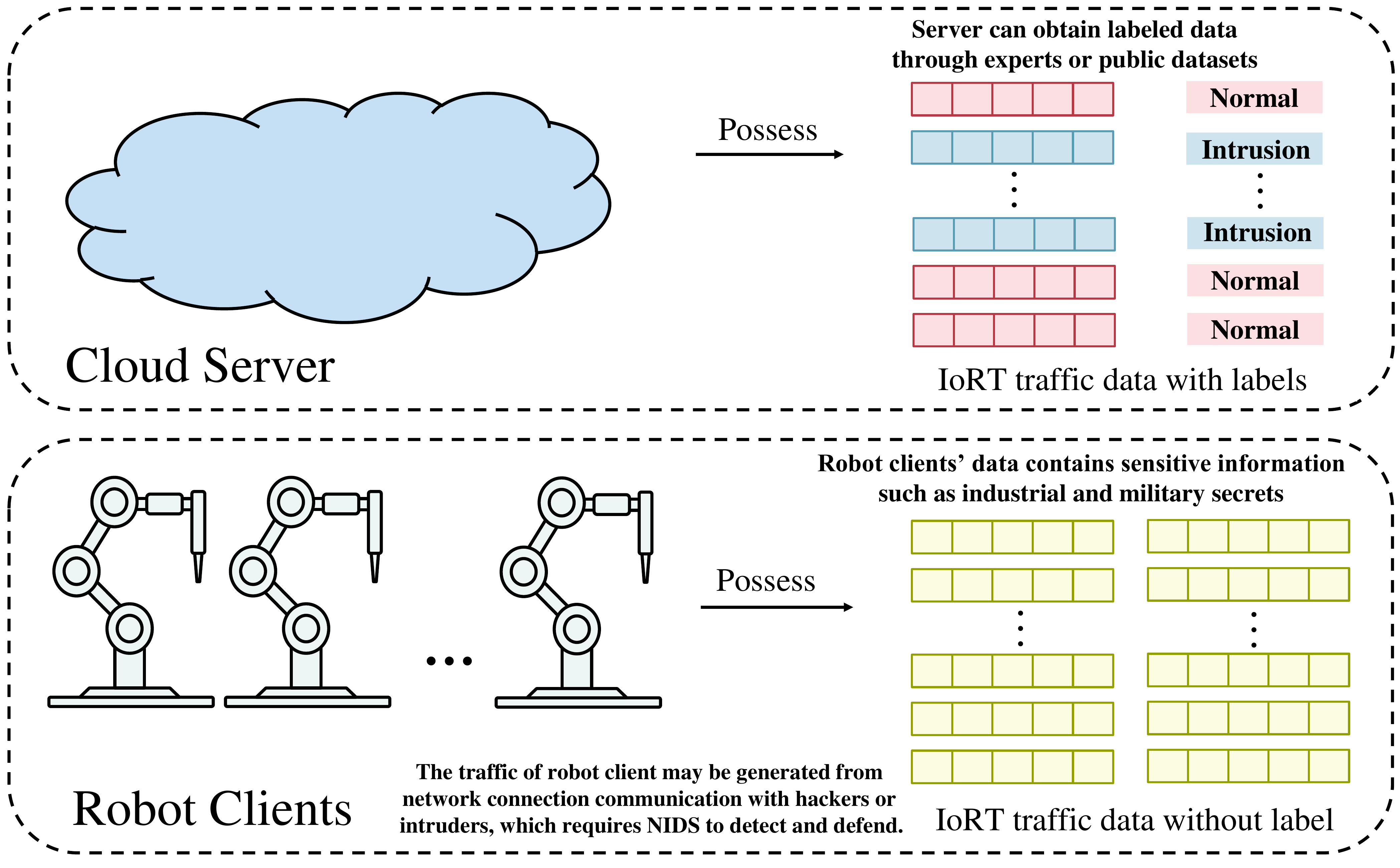}
      \caption{In pratical IoRT system, server can get labeled data by hiring experts to manually label, while robot clients do not have sufficient capability to label data.}
      \label{figurelabel}
   \end{figure}
\vspace{-5pt}
\textbf{Our solutions and contributions.} To address the aforementioned challenges in practical IoRT scenarios, we propose CFedSSL-NID tailored for IoRT. By leveraging federated learning (FL), cloud server collaboratively train IoRT DLNIDS using data and compute from decentralized robots while ensuring data remains in robot clients (privacy preserving). Semi-supervised learning (SSL) enables training even without label at robot clients, with only cloud server possessing labeled data. Robot clients employ contrastive learning (CL), a self-supervised approach that contrasts positive (similar) and negative (dissimilar) sample pairs in a latent space. Our contributions are summarized:
\begin{itemize}
    \item We implement a general approach integrating FL, SSL, and CL, capable of training with distributed unlabeled data to achieve enhancing robustness, generalization, and performance while privacy preserving.
    \item We propose CFedSSL-NID, an accurate and efficient FedSSL framework for IoRT intrusion detection with privacy preservation. It integrates random weak and strong data augmentation to boost model generalization and robustness, latent contrastive learning for performance improvement from unlabeled data, and EMA update for fine-tuning with supervised signals.
    \item We implement and validate CFedSSL-NID alongside several existing federated semi-supervised and fully supervised methods on benchmark intrusion traffic dataset. Experimental results demonstrate the effectiveness and efficiency of CFedSSL-NID.
\end{itemize}
\vspace{-5pt}
   \begin{figure}[thpb]
      \centering
      \includegraphics[width=8.5cm]{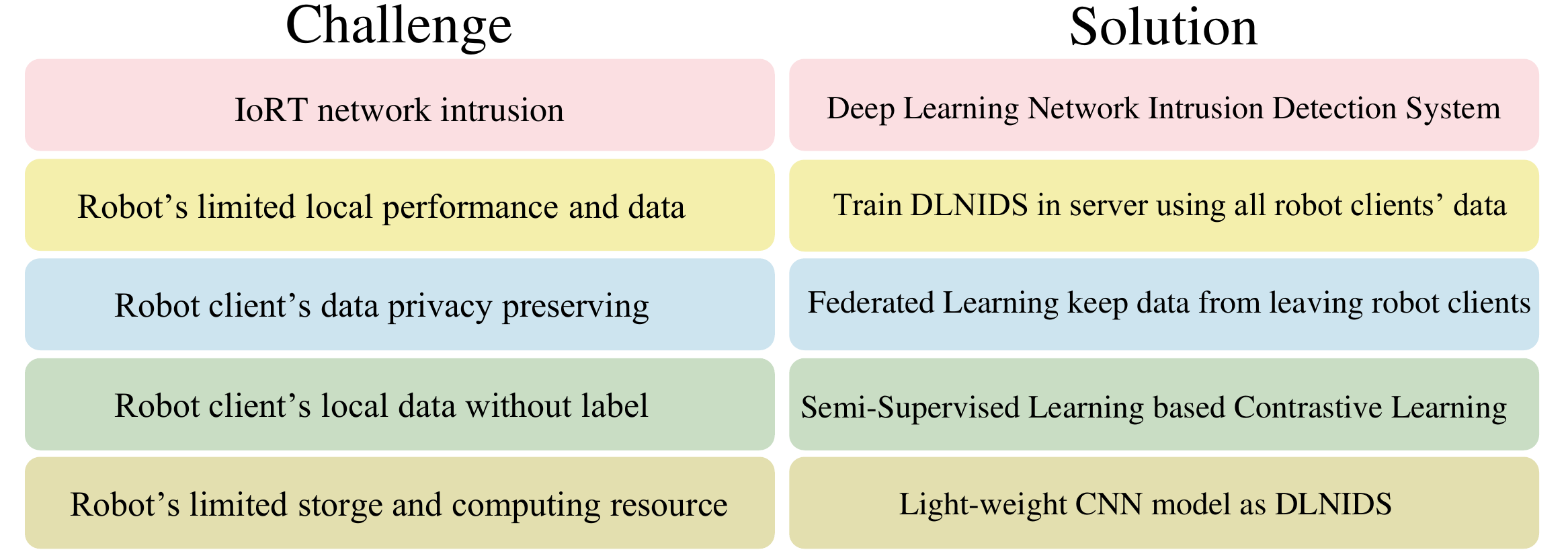}
      \caption{Brief summary of challenges in practical IoRT scenarios and solutions in our work.}
      \label{figurelabel}
   \end{figure}
\vspace{-10pt}
\section{RELATED WORKS}
\textbf{About IoRT.}
\cite{IoRT1} revisits the classification of IoRT, including its smart connectivity, architecture, and trustworthy frameworks, while investigating technologies that enhance IoRT's efficiency in executing tasks across various domains. \cite{andronie2023big} integrates IoRT's insights into big data management, deep learning object detection, and sensor fusion. \cite{mahajan2023automatic}  utilizes LSTM to construct a framework of computer vision and deep learning, enhancing the performance and efficiency of real-time IoRT applications.

\textbf{IoT intrusion detection.}
\cite{mishra2021internet} illustrates different types of DDoS and other attacks in IoT, and also explores deep learning-based intrusion detection system models. \cite{IoTSurvey} shows various IoT attacks and compares multiple machine learning models including LR, SVM, DT, RF, and ANN in IoT intrusion detection. \cite{rahman2020internet} proposed a federated learning intrusion detection scheme for IoT, which protects privacy through local training, while achieving high accuracy and low computational complexity.

\textbf{Federated semi-supervised learning.} FL addresses the challenge of training on isolated data islands, with most research focusing on fully supervised settings where every client has fully labeled data; the foundational FedAvg averages clients’ model updates on server \cite{MCMAHAN2017}. Obtaining labeled data is very expensive. SSL enhances model performance by leveraging low-cost unlabeled data, with widely adopted consistency regularization, such as UDA \cite{UDA} and Mixmatch \cite{Mixmatch}. Fixmatch \cite{fixmatch} combines pseudo-labeling with consistency regularization. Incorporating SSL methods with FL algorithms results in approaches such as FedUDA and FedFixMatch \cite{FedCy}. Methods like FedMatch \cite{FedMatch}, FedRGD \cite{FedRGD}, and FedCon \cite{FedCon} have also adopted consistency loss on FL clients. But clients only have unlabeled data causes losses to the aforementioned FedSSL method \cite{FedCon}.

\textbf{Contrastive learning.} Contrastive learning compares similar and dissimilar instances to learn discriminative representations. SimCLR leverages contrastive learning through data augmentation to learn robust visual representations from unlabeled images \cite{SimCLR}. BYOL learns data representations from unlabeled data through the interaction of two networks, without requiring negative samples for contrast \cite{BYOL}.

\section{PROPOSED METHOD}
\begin{figure*}[thpb]
   \centering
   \includegraphics[width=18cm]{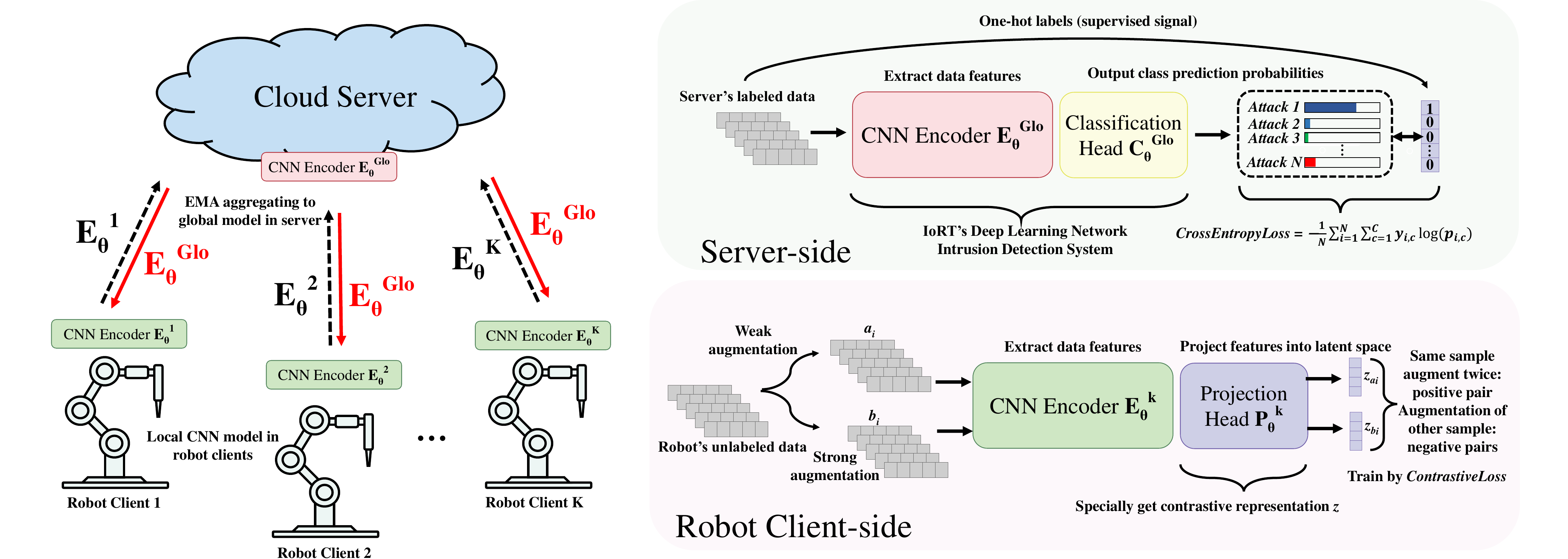}
   \caption{Overview of the proposed CFedSSL-NID framework. Each robot client k updates $E_\theta^{k}$ by local unlabeled data using $ContrastiveLoss$ and uploads $E_\theta^{k}$ to server. At server-side, global model $E_\theta^{Glo}$ will be updated by EMA with clients' model aggregation. Then $E_\theta^{Glo}$ will be updated by labeled data using $CrossEntropyLoss$ to add supervised signal.}
   \label{figurelabel}
\end{figure*}
\vspace{-5pt}

As stated in the introduction, we are more focused on a practical and realistic key feature within the Internet of Robotic Things: no labeled data in robot clients. To address this, we propose CFedSSL-NID, which utilizes contrastive learning for self-supervised learning on unlabeled data to improve model performance. The server holds labeled data $\mathcal{D}_l=\{(x_1,y_1),(x_2,y_2),...,(x_n,y_n)\}$, while robot client k possesses unlabeled data $\mathcal{D}_u^{k}=\{(x_1^{k}),(x_2^{k}),...,(x_{n_k}^{k})\}$. 
Similar to common federated learning, the server coordinates the training by aggregating model parameter updates from $K$ decentralized robot clients over $R_s$ rounds, each training local models on their respective data for $P_c$ epochs, to collaboratively improve a global model. Federated learning ensures that sensitive information stays localized and secure within the robot client devices. Overview of the proposed CFedSSL-NID is illustrated in Fig. 3. 

\subsection{Client-Side}
For each unlabeled data sample from the robot client, both weak augmentation and strong augmentation are applied. The strong/weak augmented sample pair derived from the same original sample, i.e. $x_i + \eta_{\text{weak}} = a_i$, $x_i + \eta_{\text{strong}} = b_i$. $(a_i,b_i)$ are positive pair since their semantic information remains unchanged. The augmentations of other samples, differing in semantic information from the strong/weak augmentations of this sample, are regarded as negatives. Representations $z_{a_i}$ and $z_{b_i}$ of positive pair $(a_i,b_i)$, obtained by extracting features through the CNN Encoder $E_\theta^{k}$ and projecting them into the latent space via the Projection Head $P_\theta^{k}$, should be similar. By minimizing the difference in representations of positive pairs through the $ContrastiveLoss$ and increasing the difference between representations of positive and negative samples, the CNN Encoder $E_\theta^{k}$ can learn both the difference among different samples and the commonalities among similar samples. This enables it to learn more generalized and robust features about IoRT traffic data, ultimately saving time and computational costs and significantly improving performance for downstream IoRT traffic classification and detection tasks. This achieves performance improvement utilizing unlabeled data from distributed robot clients while preserving privacy.

\textbf{Randomly weak/strong augmentation and Dropout.} Weak augmentation applies minor transformations to the original IoRT traffic data (such as adding small-scale noise), typically without altering its basic features. It may preserve the original semantic information, help the model learn basic data features and further increase the detection accuracy for downstream IoRT intrusion detection tasks \cite{FedCon}. Strong augmentation increases the diversity of training data by applying larger transformations to original IoRT traffic data, without changing its semantic information. Firstly, by introducing greater noise, strong augmentation may help the model eliminate noise interference, resulting in more stable, generalizable, and robust feature representations, which contribute to achieving better performance in downstream IoRT intrusion detection tasks. Secondly, it may improve the model's generalization ability \cite{StrongAug}. Combining weak/strong augmentations, randomly varying the augmentation scale for each batch size in every epoch, and using Dropout, can enhance data diversity and improve the generalization, stability, and robustness of feature representations \cite{Dropout}.
\vspace{-5pt}
   \begin{figure}[H]
      \centering
      \includegraphics[width=8.5cm]{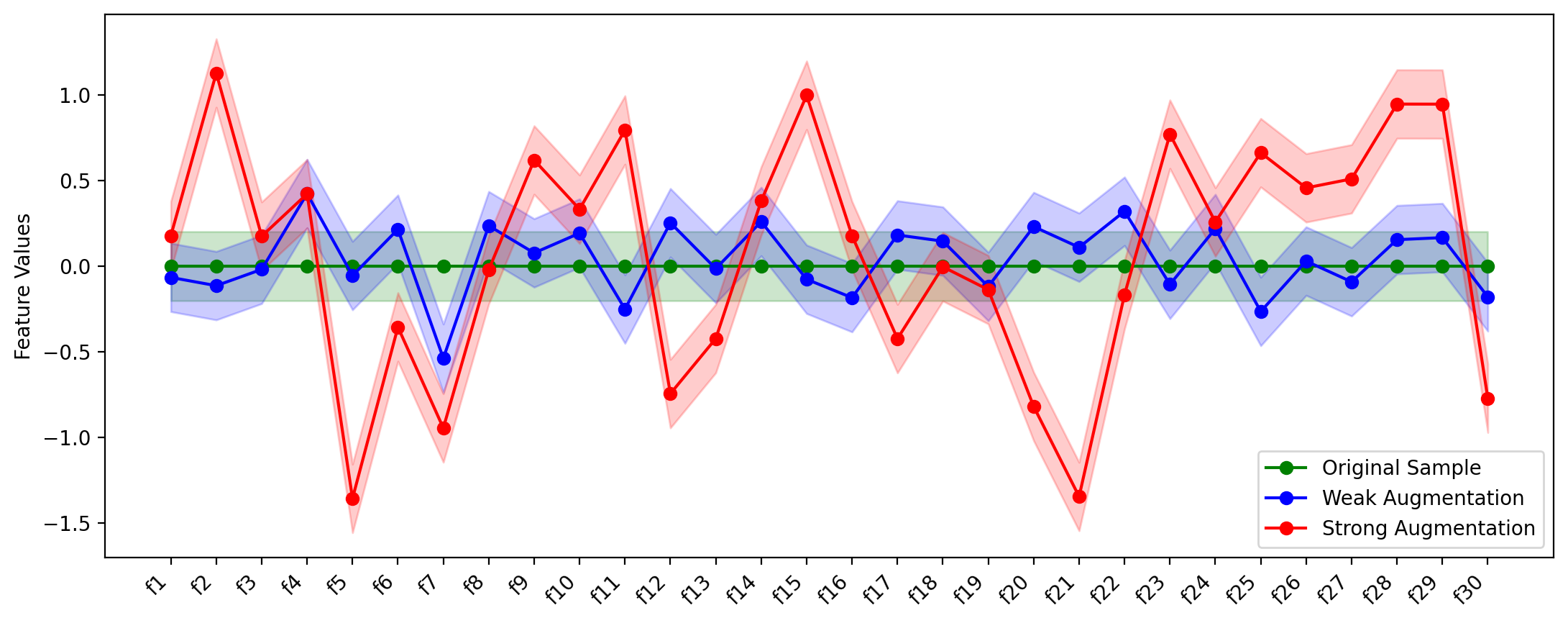}
      \caption{Example diagram comparing the numerical features of the original IoRT traffic data with its strongly and weakly augmented versions.}
      \label{figurelabel}
   \end{figure}

\textbf{Latent contrastive learning.} For the traffic data generated by IoRT (Internet of Robotic Things), we adopt a lightweight one-dimensional CNN as the encoder to extract features. Specifically, we utilize a small fully connected neural network with non-linear layers as the Projection Head $P_\theta^{k}$, which non-linearly maps the feature representations obtained from the CNN Encoder $E_\theta^{k}$ into a latent space, where contrastive loss functions are applied. Compared to directly utilizing the representations from the CNN Encoder for contrast, experimental evidence demonstrates that this approach is more effective \cite{SimCLR}. Given a positive pair of samples ($a_i, b_i$) resulting from Weak/Strong augmentations of the same instance sample $i$, we obtain their representations $z_{a_i}$ and $z_{b_i}$ in the latent space. Meanwhile, the $2(B-1)$ augmented samples from the remaining $B-1$ samples within the same batch are considered as negative samples. The normalized dot product (cosine similarity) $ \text{sim}(\textbf{a}, \textbf{b}) = \frac{\textbf{a} \cdot \textbf{b}}{\|\textbf{a}\| \|\textbf{b}\|} $ between representations is employed as the measure of similarity. The $ContrastiveLoss$ $\mathcal{L}_{\text{con}}$ can be formulated as:
\begin{equation}
 \mathcal{L}(a_i,b_i) = -\log\left(\frac{e^{\text{sim}(z_{a_i}, z_{b_i})/\tau}}{\sum_{k=1}^{B} (e^{\text{sim}(z_{a_i}, z_{a_k})/\tau} + e^{\text{sim}(z_{a_i}, z_{b_k})/\tau})}\right) 
\end{equation}
\vspace{-15pt}
\begin{equation}
 \mathcal{L}_{\text{con}} = \sum_{i=1}^{B} (\mathcal{L}(a_i,b_i) + \mathcal{L}(b_i,a_i))
\end{equation}

\vspace{-10pt}

\subsection{Server-Side}
Each robot client runs multiple epochs locally using the aforementioned contrastive method, obtaining its local parameters $\{E_\theta^{1},E_\theta^{2},E_\theta^{3},...,E_\theta^{K}\}$. These parameters are then uploaded to server, where they are aggregated using the classical FedAvg \cite{MCMAHAN2017} and get $E_\theta^{Agg} = \sum_{k=1}^{K} \frac{n_k}{n} E_\theta^{k}$. This not only ensures that data remains on the robot clients (preserving privacy), but also fully integrates and utilizes the limited computational resources of multiple robot clients for efficient training. Additionally, it enables the global model to learn different data features and distributions from each robot client, incorporating personalized knowledge from each one.

Labeled data is available in server. we employ the widely-used $CrossEntropyLoss$ $\mathcal{L}_{\text{CE}}$ for supervised learning to train the $E_\theta^{Glo}$ and $C_\theta^{Glo}$. The IoRT traffic data is processed through CNN Encoder $E_\theta^{Glo}$ to obtain representations, which are then fed into Classification Head $C_\theta^{Glo}$ to output prediction probabilities $p_{i} = C_\theta^{Glo}(E_\theta^{Glo}(x_i))$. The probability of class c \( p_{i,c} \) are compared with one-hot label \( y_{i,c} \) to calculate loss. \( N \) and \( C \) are the number of samples and classes.
\vspace{-5pt}
\begin{equation}
     \mathcal{L}_{\text{CE}} = -\frac{1}{N} \sum_{i=1}^{N} \sum_{c=1}^{C} y_{i,c} \log(p_{i,c})
\end{equation}

\textbf{Exponential moving average (EMA).} Exponential moving average updates global model $E_{\theta_{t+1}}^{Glo}$ through weighted moving averages (weight $\xi$), integrating parameters from both supervised signals (server-side $E_{\theta_t}^{Glo}$) and self-supervised learning (robot client-side $E_{\theta_{t}}^{Agg}$). As mentioned earlier, robot clients obtain stable, robust, and generalized features through contrastive learning. Before these features are applied to downstream supervised classification and detection, they need to be fine-tuned with supervised signals. This allows the generalized features to be used for class judgment. This process is achieved by obtaining new global model parameters through EMA weighting averages:
\begin{equation}  
    E_{\theta_{t+1}}^{Glo} = \xi \cdot E_{\theta_{t}}^{Glo} + (1 - \xi) \cdot E_{\theta_{t}}^{Agg}  
\end{equation}

\section{EXPERIMENTS}
We conduct extensive experiments to evaluate the proposed CFedSSL-NID including ablation and comparative studies. Experimental results show that CFedSSL-NID achieves best performance compared with baselines including federated semi-supervised and fully supervised approaches. The multi-classification performance indicators are all averaged by running more than 5 times to reduce the impact of performance fluctuations caused by randomness.

\subsection{Experimental Configuration and Metrics}
\textbf{Configuration.} Experiments were conducted on a environment with Intel (R) Xeon (R) Gold 6240 CPU @ 2.60GHz, Tesla V100S-PCIE-32GB GPU and Ubuntu 18.04.3 LTS. Code was implemented in Python 3.7.6 and PyTorch 1.13.1+cu117. Robot clients locally update 5 epochs and possess 69070 NSL-KDD unlabeled IoT traffic data. Server aggregation 10 times and possesses 50000 NSL-KDD labeled IoT traffic data. Batch size $B_S=128$ on server-side, learning rate 0.01 and Adam were utilized during training.

\textbf{Metrics.} We utilized multiple classification performance indicators to provide a comprehensive evaluation, including Accuracy (Acc), Precision (Pre), Recall, and F1-score. Given the imbalance of the data, relying solely on Acc as a metric is insufficient. Therefore, Pre, Recall, and F1-score were also used for a more thorough assessment. F1-score is particularly valuable as it considers both precision and recall, rendering it a reliable and robust metric \cite{cui2023novel}.
% \begin{equation}
% \text{Precision} = \frac{\mathbf{TP}}{\mathbf{TP} + \mathbf{FP}}
% \end{equation}
% \vspace{-5pt}
% \begin{equation}
% \text{Recall} = \frac{\mathbf{TP}}{\mathbf{TP} + \mathbf{FN}}
% \end{equation}
\vspace{-2pt}
\begin{equation}
    \text{Accuracy} = \frac{
    \mathbf{TP} + \mathbf{TN}
}{
    \mathbf{TP} + \mathbf{TN} + {\mathbf{FP}} + {\mathbf{FN}}
}
\end{equation}
\vspace{-5pt}

In multi-class scenarios, it is crucial to consider the global performance across all classes. To achieve this, calculate various classes' $\text{Pre}_i$ and $\text{Recall}_i$ separately, and use ratio of each class quantity as the weighted average:
\begin{equation}
    \text{Weighted F1} = \sum_{i=1}^{C}
\left(
    \frac{\text{quantity}_i}{\sum_{j=1}^{C} \text{quantity}_j}
    \cdot
    {2}
    \cdot
    \frac{
        {\text{Pre}_i} \cdot 
        {\text{Recall}_i}
    }{
        {\text{Pre}_i} + 
        {\text{Recall}_i}
    }
\right)
\end{equation}

Metrics for measuring complexity are Params and FLOPs.

\subsection{Dataset}
Experiments utilize the NSL-KDD~\cite{tavallaee2009detailed}, a widely used benchmark dataset in the field of IoT network intrusion detection. Many IoT intrusion detection works use NSL-KDD for evaluation \cite{IoTKDD1,IoTKDD2,IoTKDD3,IoTKDD4,IoTKDD5}. This dataset provides comprehensive and authentic IoT network intrusion traffic data, exhibiting a natural imbalance in data distribution as well as high-dimensional features. These make NSL-KDD an excellent candidate for evaluating the effectiveness and robustness of IoRT intrusion detection models. All of the following evaluations were conducted on KDDTest+.
\vspace{-5pt}
\begin{table}[h]    
\centering
\caption{NSL-KDD Description}\label{tab1}    
\begin{tabular}{|c|c|c|}    
\hline    
Class & Description & Quantity \\    
\hline    
Normal & Normal traffic without attack & 77054 \\    
DoS & Denial-of-Service: overloading to disrupt service & 53385 \\    
Probe & Information gathering by eavesdropping & 14077 \\    
R2L & Remote-to-Local: unauthorized remote access & 3749 \\    
U2R & User-to-Root: attempt to gain superuser privileges & 252 \\    
\hline    
\end{tabular}    
\end{table}
\vspace{-10pt}

\subsection{Hyperparameter Tuning and Ablation Studies}

\textbf{Hyperparameter tuning.} To tune the batch size 
$B$ of the robot clients, the temperature $\tau$ in $Contrastive Loss$, and the number of Batch Normalization (BN) in Projection Head.
\vspace{-7pt}
\begin{table}[h]  
\centering  
\caption{Hyperparameter tuning results I(\%)}  
\label{tab:example_table}  
\begin{tabular}{|c|c|c|c|c|c|c|}  
\hline  
Metrics & \multicolumn{3}{c|}{$B$=1024, $\tau$=1} & \multicolumn{3}{c|}{$B$=1024, BN=0} \\ \cline{2-7}  
& \textbf{BN=0} & BN=1 & BN=2 & $\tau$=0.07 & \textbf{$\tau$=0.5} & $\tau$=1 \\ \hline  
Acc & \textbf{79.34} & 79.09 & 78.19 & 78.33 & \textbf{80.82} & 79.34 \\ \hline  
Pre & \textbf{81.47} & 81.65 & 79.15 & 78.09 & \textbf{82.63} & 81.47 \\ \hline  
Recall & \textbf{79.34} & 79.09 & 78.19 & 78.33 & \textbf{80.82} & 79.34 \\ \hline  
F1 & \textbf{76.93} & 76.06 & 75.29 & 75.35 & \textbf{79.20} & 76.93 \\ \hline  
\end{tabular}  
\end{table}
\vspace{-15pt}
\begin{table}[h]  
\centering  
\caption{Hyperparameter tuning results II(\%)}  
\label{tab:example_table_other_conditions}  
\begin{tabular}{|c|c|c|c|c|c|c|}  
\hline  
\multicolumn{1}{|c|}{Metrics} & \multicolumn{6}{c|}{BN=0, $\tau$=0.5} \\ \cline{2-7}  
\multicolumn{1}{|c|}{} & \multicolumn{1}{c|}{$B$=128} & \multicolumn{1}{c|}{$B$=256} & \multicolumn{1}{c|}{$B$=512} &\multicolumn{1}{c|}{\textbf{$B$=1024}}& \multicolumn{1}{c|}{$B$=2048} &\multicolumn{1}{c|}{$B$=4096}\\ \hline  
Acc & 77.09  & 76.95& {77.94} & \textbf{80.82} & 76.84 & 76.02  \\ \hline  
Pre &  77.30  & 77.65 & {78.65} & \textbf{82.63}& 75.84 & 77.11\\ \hline  
Recall & 77.09  & 76.95 & {77.94} & \textbf{80.82} & 76.84 & 76.02\\ \hline  
F1 & 74.51  & 74.40 & {75.59} & \textbf{79.20}& 73.49 & 72.95\\ \hline   
\end{tabular}  
\end{table}

Applying batch normalization (BN) may degrade the stability and consistency of representations, thereby affecting the effectiveness of contrastive learning. And the setting of temperature $\tau$ of $ContrastiveLoss$ is crucial as it enables a precise balance between enhancing the model's focus on hard negative samples and maintaining the uniformity of the feature space \cite{SimCLR}. Moreover, a moderate batch size $B=1024$ may enhance the abundance of negative samples \cite{MoCo}, while optimizing computational efficiency and memory usage especially for robot clients with limited hardware.

\textbf{Ablation studies.} We conduct ablation studies on the strategies of randomly weak/strong augmentation and Dropout, robot client-side latent contrastive learning, as well as server-side EMA update, to validate their effectiveness in federated learning for IoRT intrusion detection.
\begin{table}[H]      
\centering      
\caption{Ablation studies results for CFedSSL-NID(\%)}    
\label{tab4.8}      
\begin{tabular}{|c|c|c|c|c|}      
\hline      
Methods & Acc & Pre & Recall & F1 \\ \hline
$w/o$ W/S Augs and Dropout & 77.35 & 79.31 & 77.35 & 74.81 \\  
\hline 
$w/o$ Latent Contrastive & 76.67 & 77.28 & 76.67 & 73.12 \\   
\hline 
$w/o$ EMA Update & 78.37 & 80.47 & 78.37 & 75.23 \\   
\hline 
\textbf{CFedSSL-NID} & \textbf{80.82} & \textbf{82.63} & \textbf{80.82} & \textbf{79.20 } \\  
\hline      
\end{tabular}      
\end{table} 
Contrastive learning on robot client-side significantly boost performance from unlabeled data, achieving an improvement of 4.15\% in Acc, 6.08\% in F1, and 5.35\% in Pre. Weak/strong augmentations and Dropout enhances performance by introducing randomness and diversity in data augmentation. Furthermore, the EMA update maintains the stability and effectively integrates the capabilities learned from both supervised and unsupervised learning, leading to notable performance improvements.

\textbf{Visualizations.} We visualized several above strategies through additional experiments and attempted to intuitively demonstrate their effects. The strong/weak augmentations noises and augmented data were compared with the original data (one NSL-KDD IoT traffic) for visualization. By utilizing t-SNE \cite{tsne} dimensionality reduction visualization, we compared the projection representations learned with and without Dropout. The representations learned with Dropout could be roughly observed to be divided into five clusters, demonstrating the remarkable ability to learn approximate class differences from the data itself without the use of class supervision labels, thus providing separable and generalized feature representations for downstream classification tasks.

\begin{figure}[htb]
    % 两行两列的图片布局
    \centering
    \begin{minipage}[b]{\linewidth} % 第一行，占页面宽度
        \centering
        \begin{minipage}{0.4\linewidth} % 第一列
            \includegraphics[width=\linewidth]{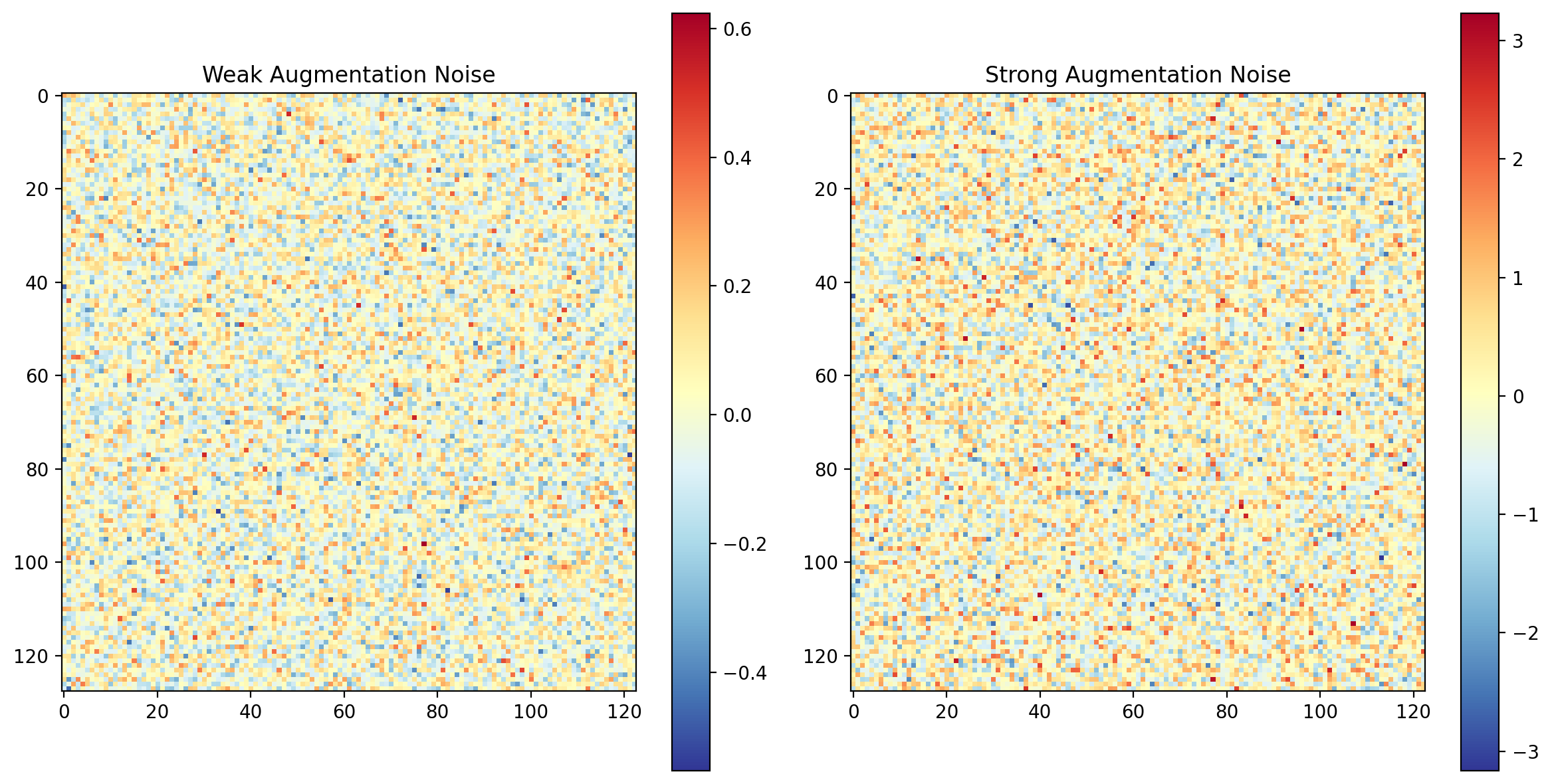}
            \centerline{(a)}
            % \label{fig:scenarios-acc}
        \end{minipage}
        % \hfill % 填充空白以分隔图片
        \begin{minipage}{0.55\linewidth} % 第二列
            \includegraphics[width=\linewidth]{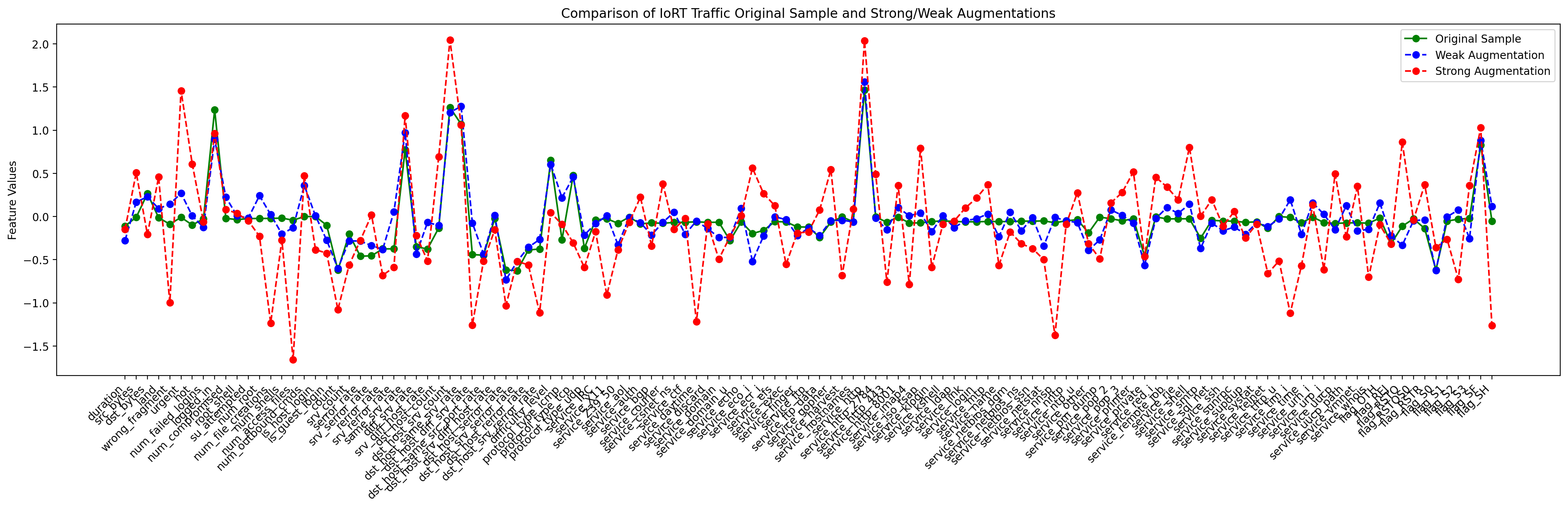}
            \centerline{(b)}
            % \label{fig:scenarios-f1}
        \end{minipage}
   \end{minipage}
    \vspace{-2pt}
    \caption{Visualizations of strong/weak augmentations noises and augmented data and original data.}
    \label{fig:performance}
\end{figure}

\vspace{-20pt}

\begin{figure}[htb]
    % 两行两列的图片布局
    \centering
    \begin{minipage}[b]{\linewidth} % 第一行，占页面宽度
        \centering
        \begin{minipage}{0.49\linewidth} % 第一列
            \includegraphics[width=\linewidth]{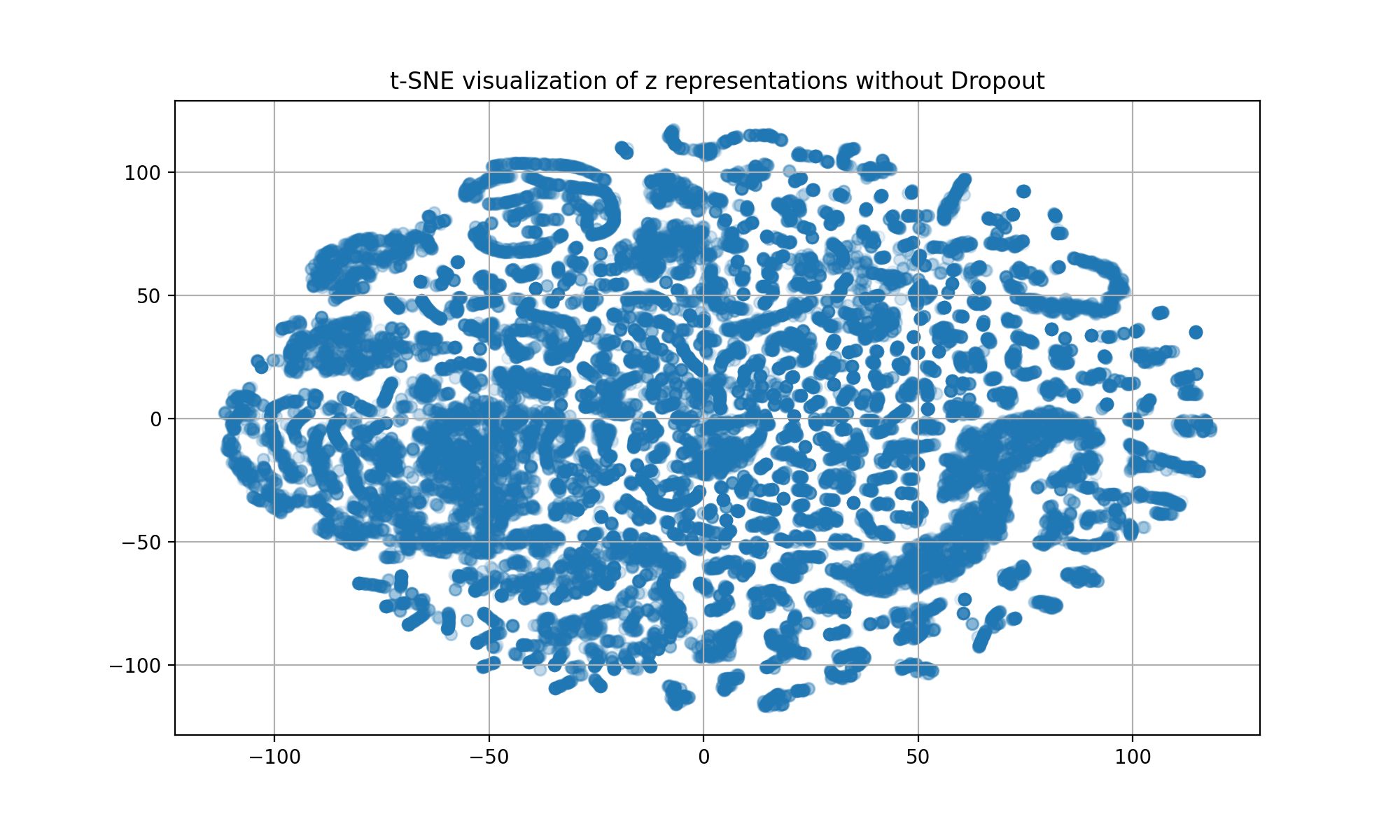}
            \centerline{(a) Without Dropout}
            % \label{fig:scenarios-acc}
        \end{minipage}
        % \hfill % 填充空白以分隔图片
        \begin{minipage}{0.49\linewidth} % 第二列
            \includegraphics[width=\linewidth]{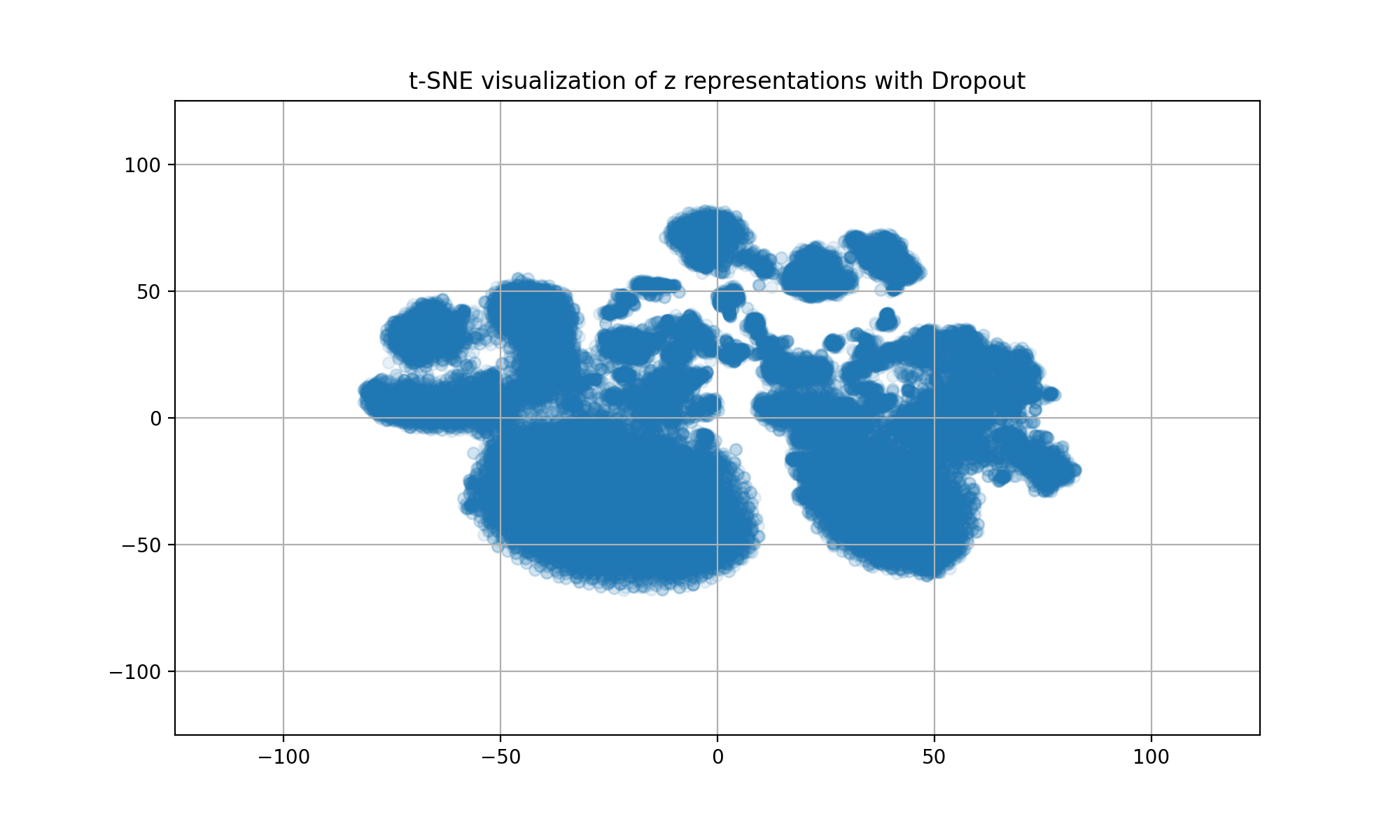}
            \centerline{(b) With Dropout}
            % \label{fig:scenarios-f1}
        \end{minipage}
   \end{minipage}
    \vspace{-2pt}
    \caption{T-SNE visualizations to compared the projection representations learned with and without Dropout.}
    \label{fig:performance}
\end{figure}

\vspace{-15pt}

\subsection{Comparative Experiments}
The baselines for the detection performance comparison experiments uniformly employ lightweight CNNs as both global and local models. Federated semi-supervised learning: integrate semi-supervised methods Fixmatch \cite{fixmatch}, UDA \cite{UDA}, and CR (Consistency Regularization, where clients apply mean square error loss and cosine similarity loss for consistency regularization on model representations of augmented data pairs) \cite{CR} with federated algorithms FedAvg \cite{MCMAHAN2017} and FedProx \cite{FedProx}. Federated supervised methods: SFedAvg\_AD \cite{MCMAHAN2017} (Supervised FedAvg with All 125973 training traffic Data from NSL-KDD evenly distributed across 10 clients) and SFedProx\_AD \cite{FedProx} (Supervised FedProx using All Data). Fully supervised centralized learning: CSL\_SD (Centralized Supervised Learning by the Server's 50000 Data) and CSL\_AD (Centralized Supervised Learning utilizing the All 125973 NSL-KDD Data). It must be noted that the original setting of federated semi-supervised learning baselines assumes that clients have some labeled data, which is clearly inconsistent with the actual scenarios of robot clients in IoRT. In our experiments, only these baselines' self-supervised methods on unlabeled data were adopted. This underscores the CFedSSL-NID's tailored design for the practical scenarios of robot clients in IoRT, which differs from previous methods.

\textbf{Binary classification comparison.} For IoRT intrusion traffic,  we can simply divide into two categories: attack and normal. In this case, when the IoRT network intrusion detection system equipped in robot identifies IoRT traffic in the attack category, it can send an alert to remind the administrator or automatically take measures against abnormal traffic connections according to program. We provide comparison of binary classification confusion matrix (Fig. 7) and the performance indicators calculated from it.

\begin{figure}[H]
    \centering
    \begin{minipage}[b]{\linewidth} 
        \centering
        \begin{minipage}[b]{0.18\linewidth} 
            \includegraphics[width=\linewidth]{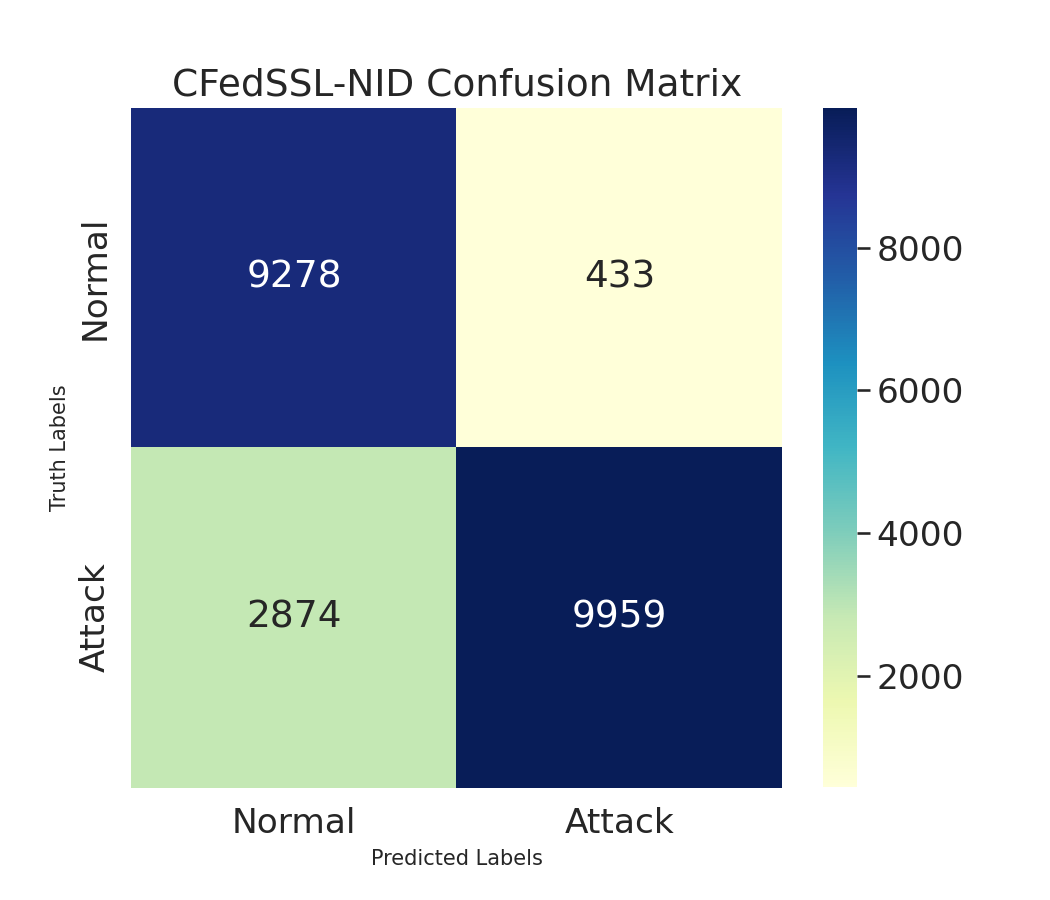}
            \centerline{(a)}
            \label{fig:s1-acc}
        \end{minipage}
        \hfill 
        \begin{minipage}[b]{0.18\linewidth} % 第二列
            \includegraphics[width=\linewidth]{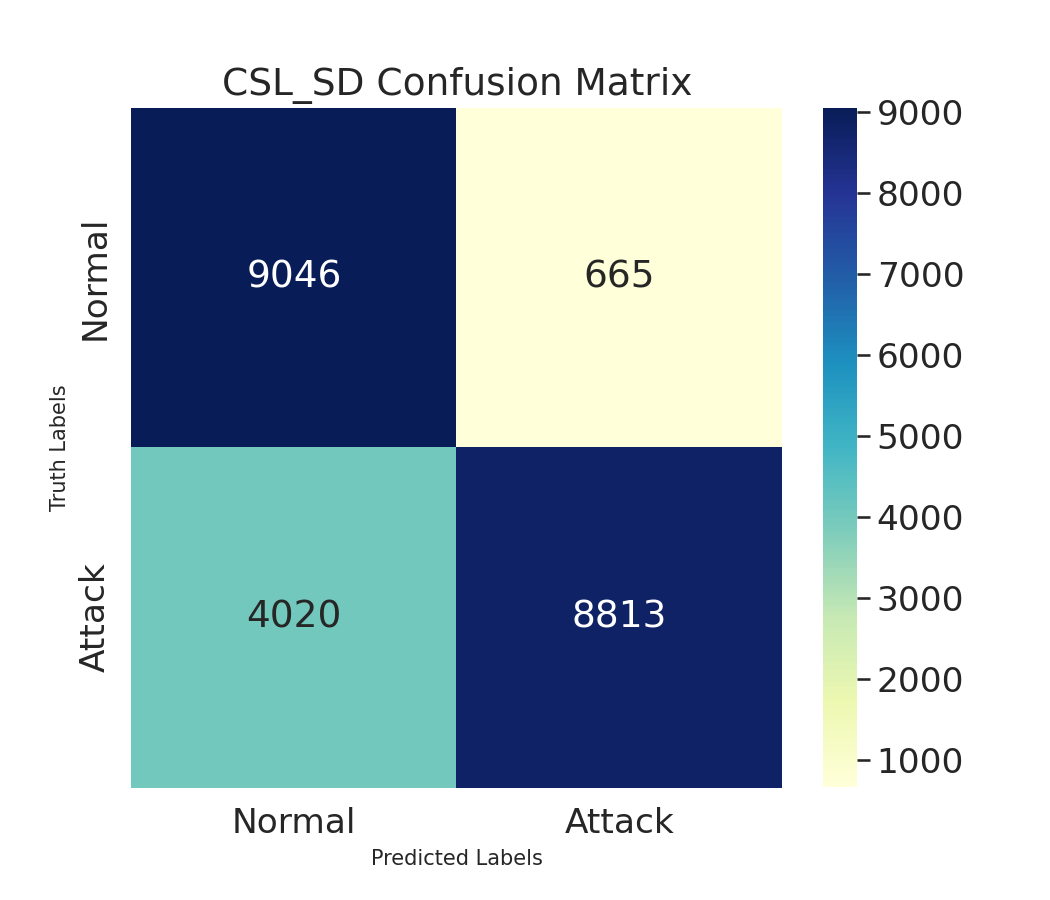}
            \centerline{(b)}
            \label{fig:s2-acc}
        \end{minipage}
        \hfill
        \begin{minipage}[b]{0.18\linewidth} % 第三列
            \includegraphics[width=\linewidth]{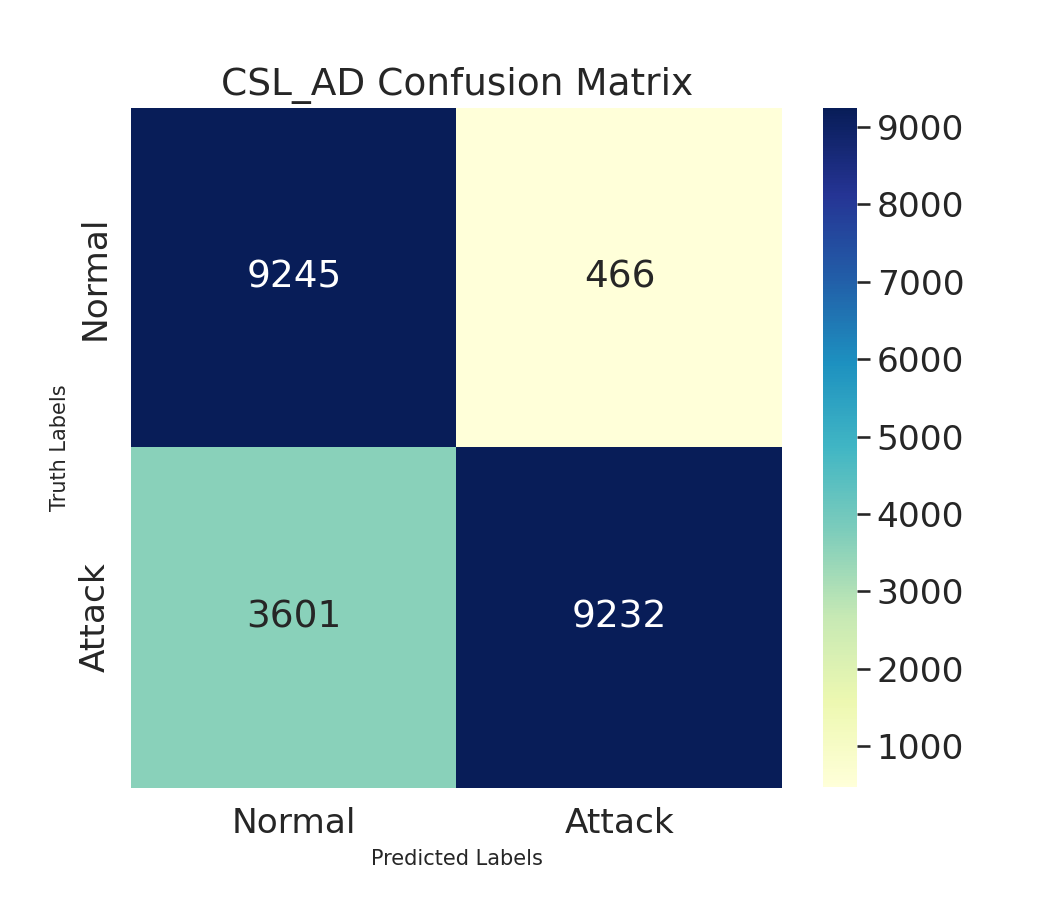}
            \centerline{(c)}
            \label{fig:s3-acc}
        \end{minipage}
        \hfill
        \begin{minipage}[b]{0.18\linewidth} 
            \includegraphics[width=\linewidth]{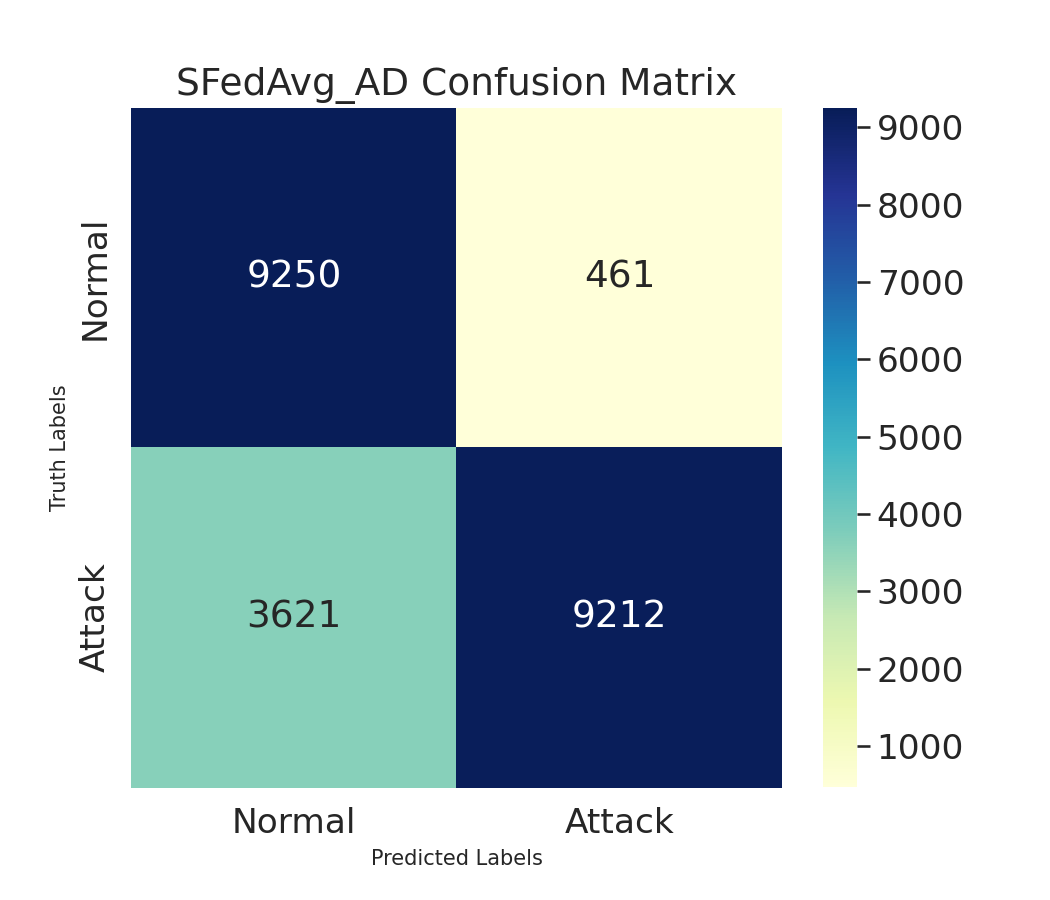}
            \centerline{(d)}
            \label{fig:s3-acc}
        \end{minipage}
                \hfill
        \begin{minipage}[b]{0.18\linewidth} 
            \includegraphics[width=\linewidth]{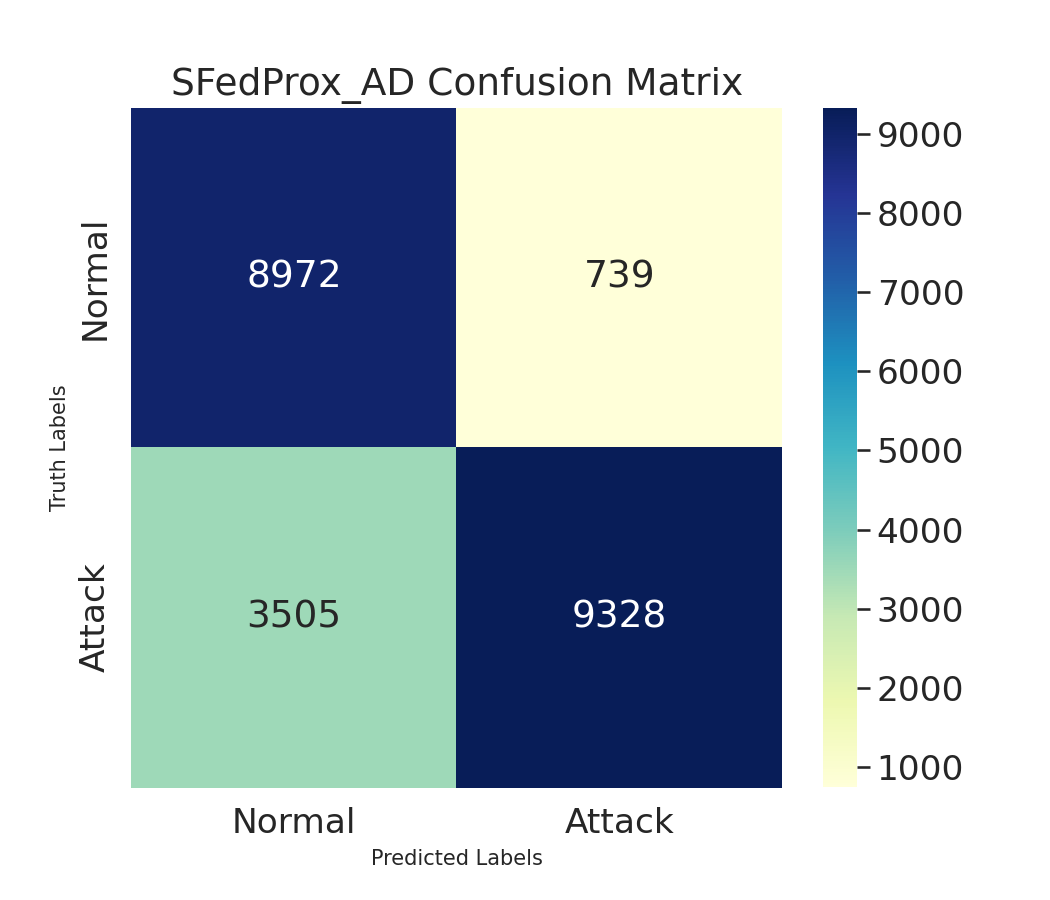}
            \centerline{(e)}
            \label{fig:s3-acc}
        \end{minipage}
    \end{minipage}
    \vspace{-20pt}
    \bigskip % 较大的间距，分隔两行
    \begin{minipage}[b]{\linewidth} 
        \centering
        \begin{minipage}[b]{0.18\linewidth} 
            \includegraphics[width=\linewidth]{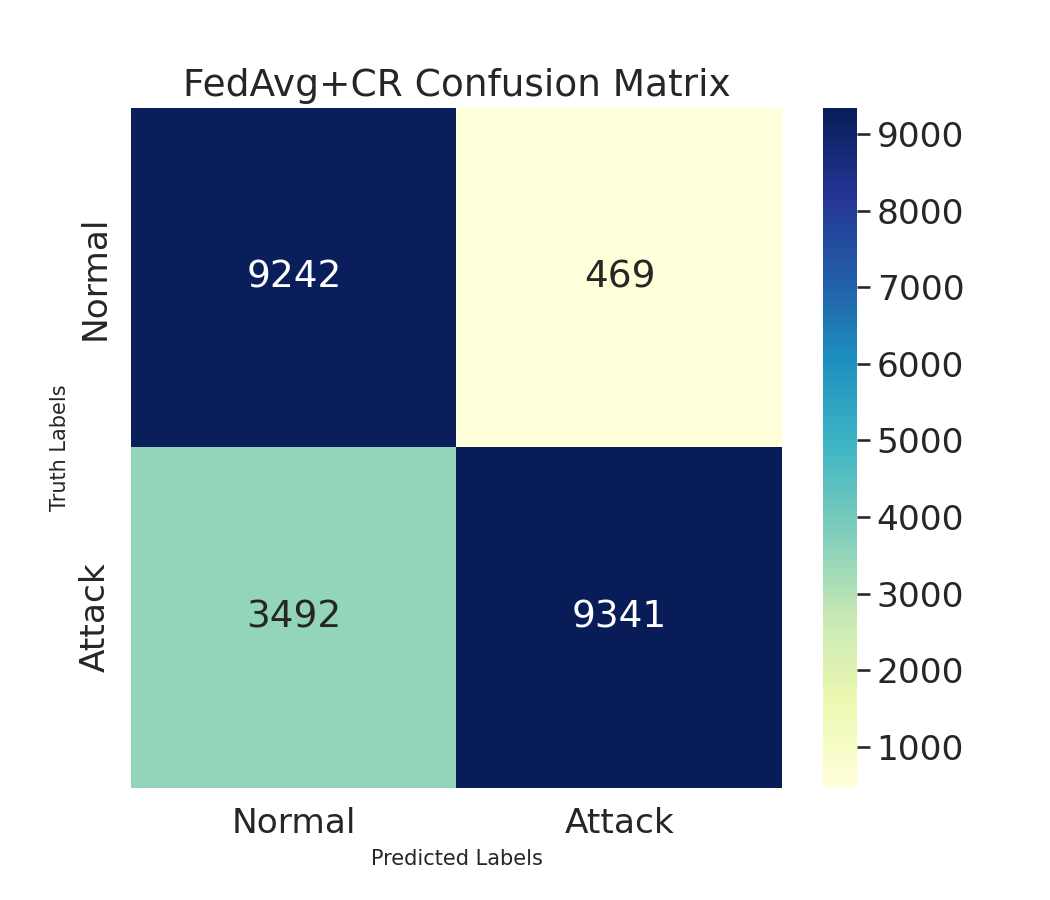}
            \centerline{(f)}
            \label{fig:s1-acc}
        \end{minipage}
        \hfill % 填充空白以分隔图片
        \begin{minipage}[b]{0.18\linewidth} % 第二列
            \includegraphics[width=\linewidth]{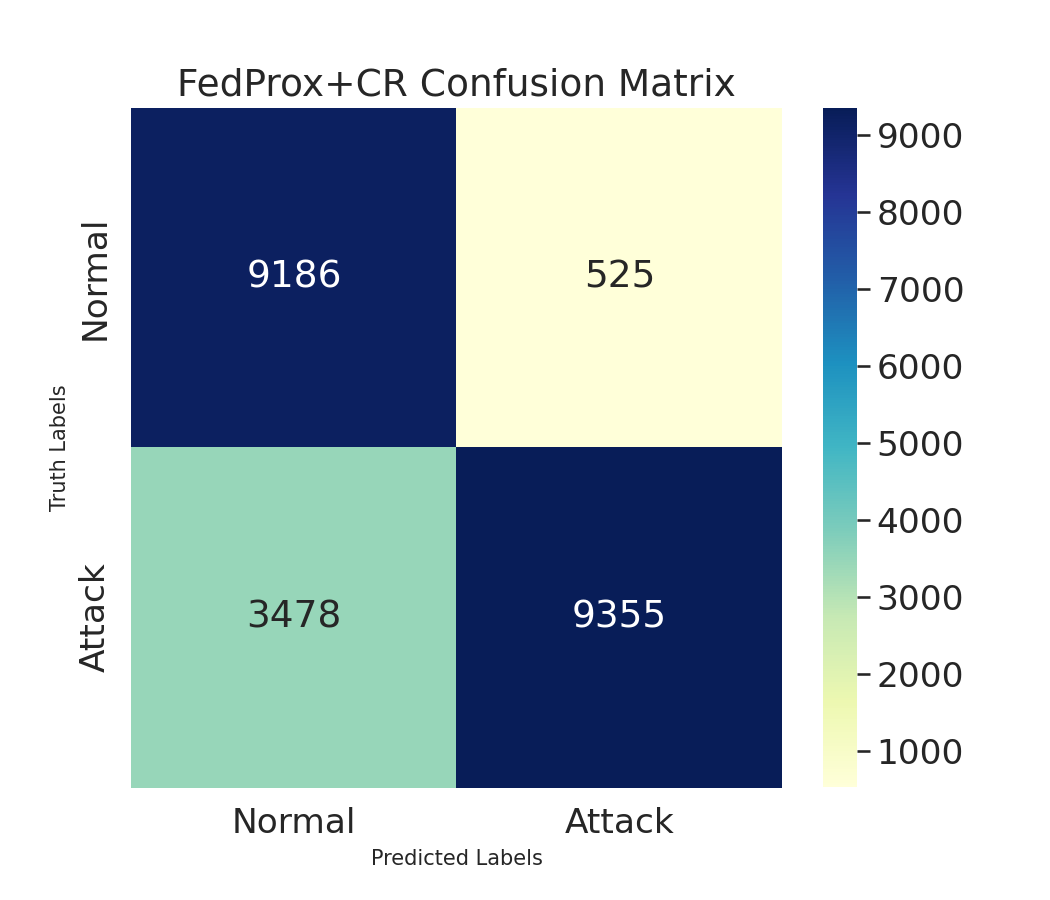}
            \centerline{(g)}
            \label{fig:s2-acc}
        \end{minipage}
        \hfill
        \begin{minipage}[b]{0.18\linewidth} % 第三列
            \includegraphics[width=\linewidth]{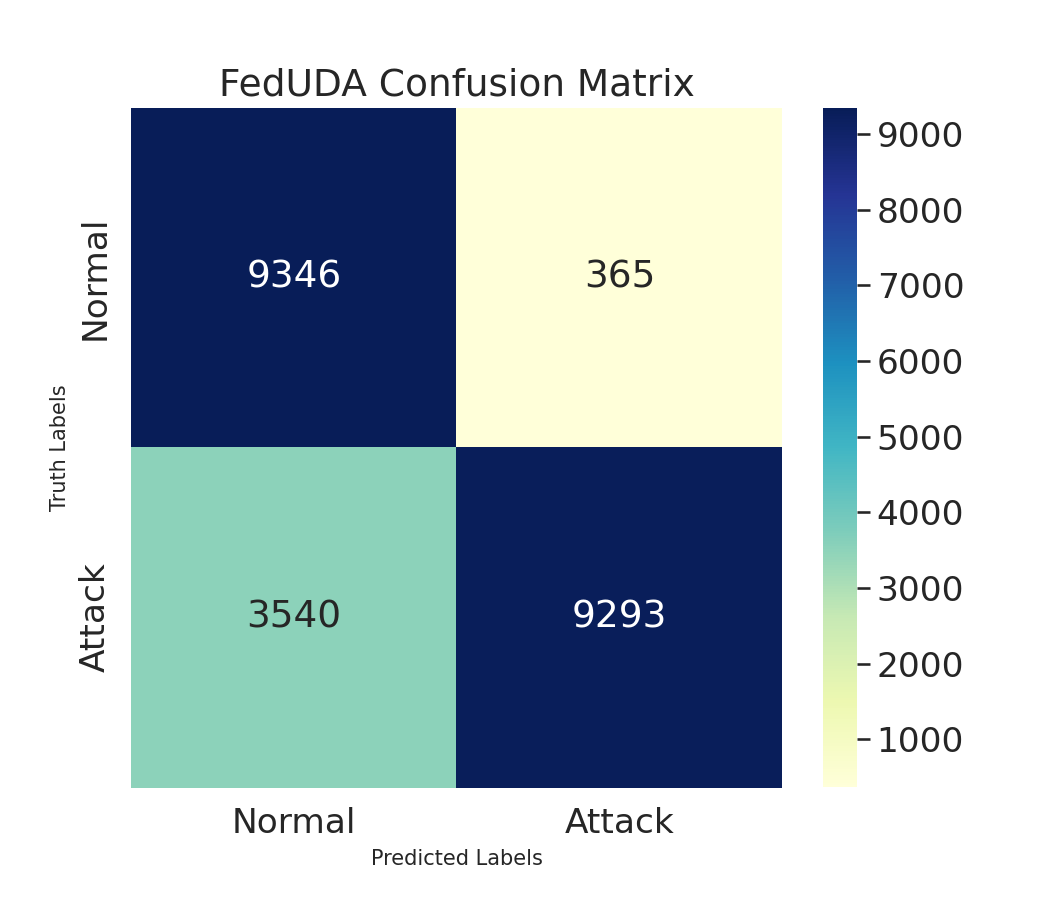}
            \centerline{(h)}
            \label{fig:s3-acc}
        \end{minipage}
        \hfill
        \begin{minipage}[b]{0.18\linewidth} 
            \includegraphics[width=\linewidth]{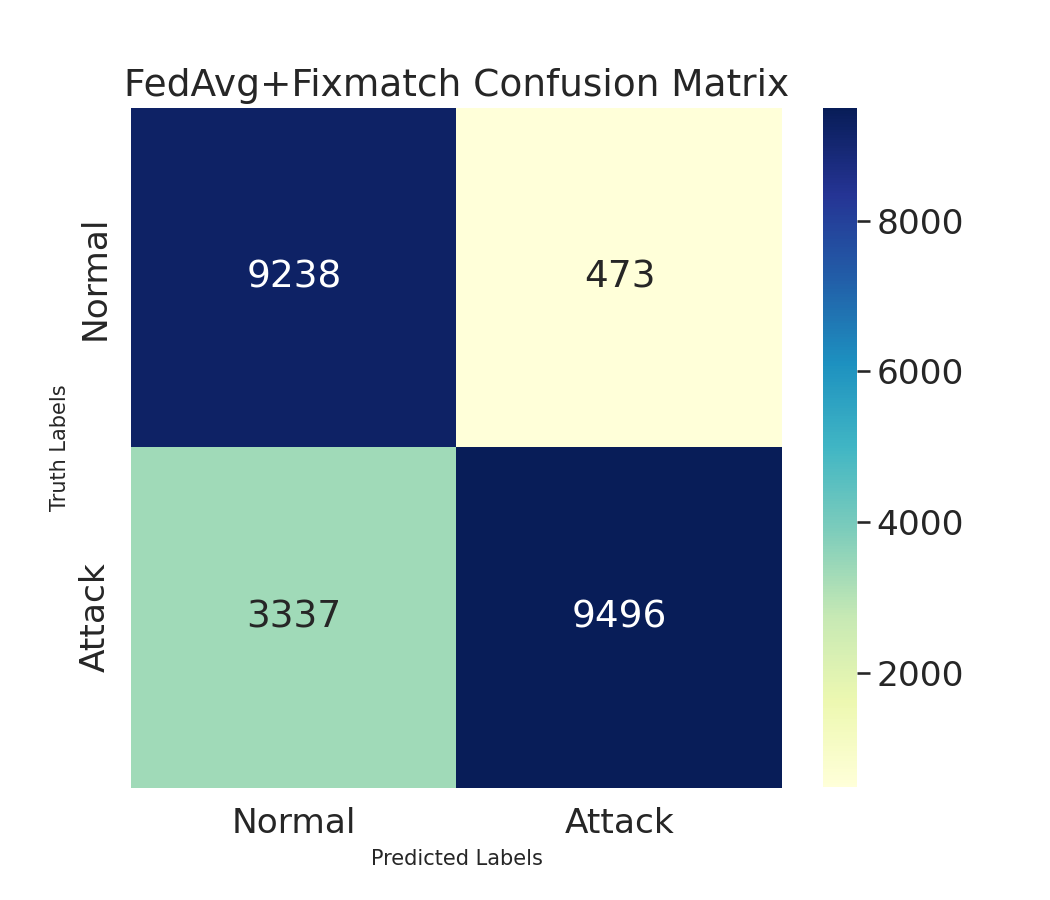}
            \centerline{(i)}
            \label{fig:s3-acc}
        \end{minipage}
                \hfill
        \begin{minipage}[b]{0.18\linewidth} 
            \includegraphics[width=\linewidth]{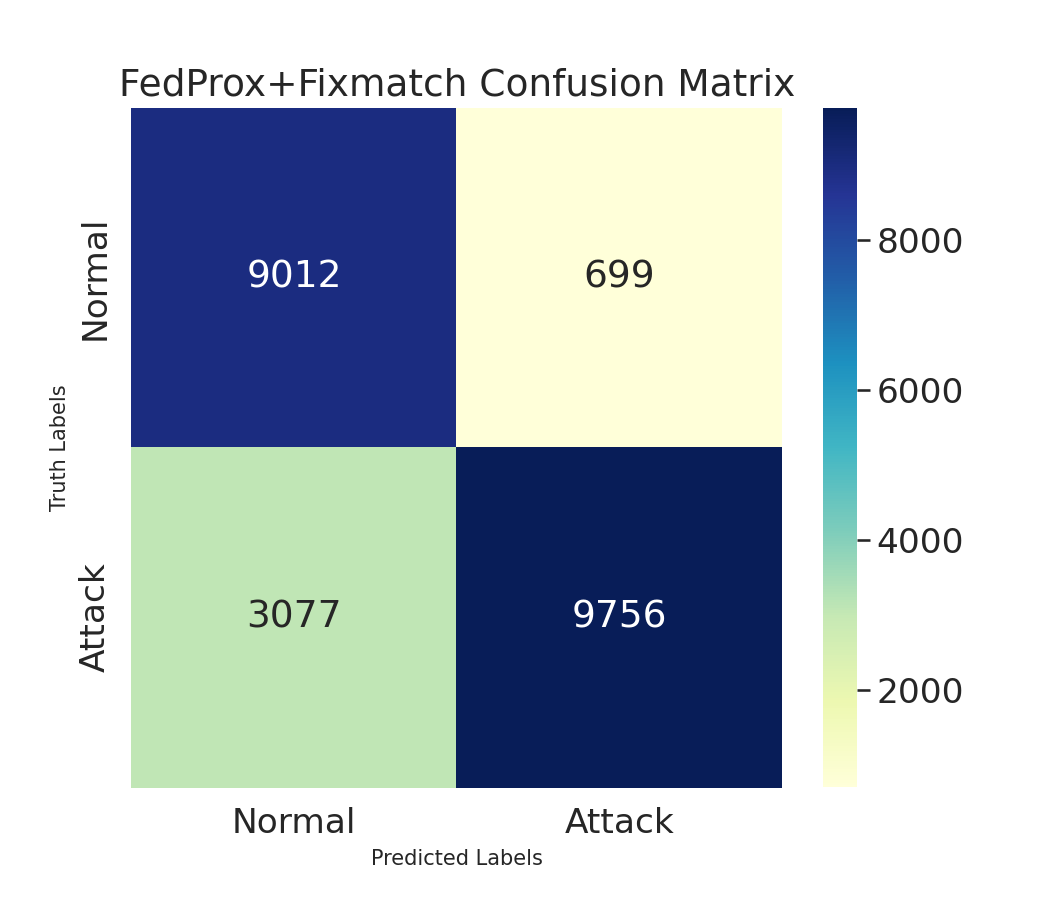}
            \centerline{(j)}
            \label{fig:s3-acc}
        \end{minipage}
    \end{minipage}
    \vspace{-15pt}
    \caption{Binary confusion matrices: (a) CFedSSL-NID (b) SFedAvg\_AD (c) SFedProx\_AD (d) CSL\_SD (e) CSL\_AD (f) FedAvg+CR (g) FedProx+CR (h) FedUDA (i) FedAvg+Fixmatch (j) FedProx+Fixmatch}
    \label{fig:comparative-experiments}
\end{figure}

\vspace{-15pt}

\begin{table}[H]      
\centering      
\caption{Binary classification comparative results(\%)}          
\begin{tabular}{|c|c|c|c|c|}      
\hline      
{Frameworks} & {Acc} & {Pre} & {Recall} & {F1} \\   
\hline 
SFedAvg\_AD \cite{MCMAHAN2017} & 81.89 & {95.23} & 71.78 & 81.86 \\  \hline
SFedProx\_AD \cite{FedProx} & 81.17 & 92.66 & 72.69 & {81.47} \\  \hline
CSL\_SD & 79.22 & 92.98 & {68.67} & 79.00 \\  \hline
CSL\_AD  & 81.96 & 95.19 & 71.94 & 81.95 \\  \hline
FedAvg+CR \cite{MCMAHAN2017,CR} &82.43& 95.22 & 72.79 & 82.51 \\  \hline
FedProx+CR \cite{FedProx,CR} & 82.24 & 94.69 & 72.90 & 82.38 \\  \hline
FedUDA \cite{MCMAHAN2017,UDA} & 82.68 & \textbf{96.22} & 72.41 & 82.64 \\  \hline
FedAvg+Fixmatch \cite{MCMAHAN2017,fixmatch} & 83.10 & 95.26 & 74.00 & 82.39 \\  \hline
FedProx+Fixmatch \cite{FedProx,fixmatch} & 83.25 & 93.31 & 76.02 & 83.79 \\  \hline
\textbf{CFedSSL-NID} & \textbf{85.33} & {95.83} & \textbf{77.60} & \textbf{85.76} \\   
\hline      
\end{tabular}      
\end{table}

The binary classification comparative experiments  demonstrate that the proposed CFedSSL-NID outperforms even the best baseline, specifically, an 2.08\% increase in Acc and 1.97\% in F1 Score. These indicators are derived from the average results of over 5 repeated runs, which minimizes the impact of randomness.

\textbf{Multi-classification comparison.} By accurately classifying the attack traffic into more detailed and specific categories (DoS, Probe, R2L, U2R in NSL-KDD), the IoRT DLNIDS can take more targeted measures to defend against intrusions. By presenting the comparison of detection performance in a multi-class classification scenario in TABLE VI, it can observe that the proposed CFedSSL-NID outperforms these federated semi-supervised, fully supervised and centralized supervised methods. 
\vspace{-5pt}
\begin{table}[H]      
\centering      
\caption{Multi-classification comparative results(\%)}          
\begin{tabular}{|c|c|c|c|c|}      
\hline      
{Frameworks} & {Acc} & {Pre} & {Recall} & {F1} \\   
\hline 
SFedAvg\_AD \cite{MCMAHAN2017} & 78.25 & {79.33} & 78.25 & 74.97 \\  \hline
SFedProx\_AD \cite{FedProx} & 77.24 & 78.29 & 77.24 & {73.90} \\  \hline
CSL\_SD & 76.67 & 77.28 & {76.67} & 73.12 \\  \hline
CSL\_AD  & 78.84 & 81.11 & 78.84 & 74.93 \\  \hline
FedAvg+CR \cite{MCMAHAN2017,CR} & 78.21 & 79.20 & 78.21 & 75.57 \\  \hline
FedProx+CR \cite{FedProx,CR} & 79.41 & 81.02 & 79.41 & 77.14 \\  \hline
FedUDA \cite{MCMAHAN2017,UDA} & 78.73 & 80.41 & 78.73 & 75.59 \\  \hline
FedAvg+Fixmatch \cite{MCMAHAN2017,fixmatch} & 79.56 & 80.51 & 79.56 & 77.32 \\  \hline
FedProx+Fixmatch \cite{FedProx,fixmatch} & 79.03 & 81.04 & 79.03 & 77.19 \\  \hline
\textbf{CFedSSL-NID} & \textbf{80.82} & \textbf{82.63} & \textbf{80.82} & \textbf{79.20} \\   
\hline      
\end{tabular}      
\end{table}

\vspace{-5pt}
The IoRT intrusion traffic in real world exhibits notable imbalance, meaning that the volume of traffic from certain categories is abundant, while others are few. Consequently, models often struggle to achieve well performance on minority classes and develop a bias towards majority classes. In NSL-KDD, Probe, R2L, and U2R are minority classes. As evident from the multi-class confusion matrix (Fig. 8) and performance metrics for each class (TABLE VII), CFedSSL-NID is still capable of achieving well detection performance on minority classes, thereby have well overall performance on imbalanced IoRT traffic data.

\vspace{-5pt}

\begin{table}[H]      
\centering      
\caption{CFedSSL-NID performances on each class(\%)}
\begin{tabular}{|c|c|c|c|c|c|}      
\hline      
{Classes} & Imbalanced Ratio & {Pre} & {Recall} & {F1} \\   
\hline 
Normal & 1.00 & {76.35} & 95.54 & 84.87 \\  \hline
DoS & 1.44 & 93.70 & 80.80 & {86.77} \\  \hline
Probe & 5.47 & 69.70 & {83.31} & 75.90 \\  \hline
R2L & 20.55 & 90.05 & 30.20 & 45.23 \\  \hline
U2R & 305.77 & 46.94 & 11.50 & 18.47 \\  \hline     
\end{tabular}      
\end{table}

\vspace{-20pt}

\begin{figure}[H]
      \centering
      \includegraphics[width=9.0cm]{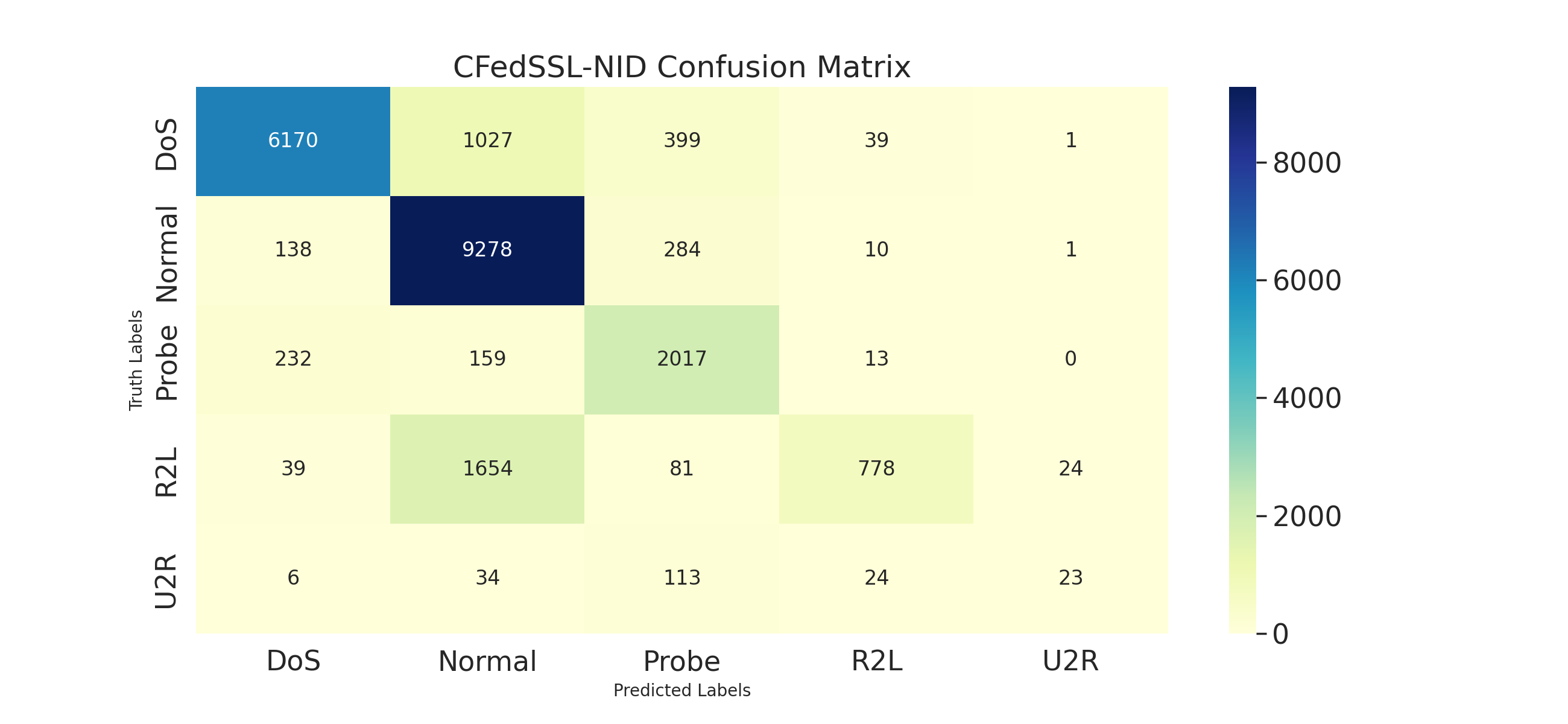}
      \caption{CFedSSL-NID multi-classification confusion matrix.}
      \label{figurelabel}
   \end{figure}

\vspace{-10pt}
\textbf{Complexity comparison.} IoRT NIDS equipped in robotic devices should be lightweight and real-time. CFedSSL-NID employs a lightweight CNN (lw-CNN) as both global and local model for server and robot clients.

% We compare the Params and FLOPs of lw-CNN with other deep learning models for DLNIDS including including Ding's CNN \cite{DingCNN}, DNN 2, 3, 4 layers, 1D-CNN \cite{Params4}, 2D CNN \cite{2dcnn}, IBYOL-IDS, CMAE, SS-Deep-ID, E-GraphSAGE, LSTM-FCNN, RNN \cite{lightweigh}, Transformer-CNN-LSTM \cite{Transformer}, PyConv \cite{PyConv}.

\vspace{-5pt}

% \begin{table}[H]      
% \centering      
% \caption{Comparison on complexity of DLNIDS models}  
% \resizebox{8.7cm}{!}{
% \begin{tabular}{|c|c|c|c|c|}  
% \hline 
% Indicator  & Ding's CNN & DNN 2layers & DNN 3layers  & 1D-CNN \\  
% \hline 
%  Params  & 126826 & 841221 & 1235717  & 90373 \\  
% FLOPs  & - & 1680670 & 2469150   & 6886280\\ \hline
% Indicator & DNN 4layers & 2D CNN & IBYOL-IDS & CMAE  \\ \hline
% Params & 1366789 & 1966086 & 1578145 & 130577 \\  
% FLOPs & 2731038 & - & - & - \\  \hline
% Indicator & SS-Deep-ID & E-GraphSAGE & LSTM-FCNN & RNN  \\ \hline
% Params & 663434 & 94722 & 1137157 & 129157 \\  
% FLOPs & - & - & - & 133300224 \\ \hline
% Indicator & Transformer-CNN-LSTM  & PyConv & \textbf{lw-CNN} &  \\ \hline
% Params & 68166 & 55877 & \textbf{49469} &  \\  
% FLOPs & - & - &\textbf{729000} &\\ 
% \hline      
% \end{tabular} 
% }
% \end{table}

\begin{table}[H]      
    \centering      
    \caption{Comparison on complexity of DLNIDS models}  
    \resizebox{8.0cm}{!}{
    \begin{tabular}{|c|c|c|c|c|}
        \hline 
        Model & Params & FLOPs & \textbf{Cite} \\  
        \hline 
        Ding's CNN  & 126826 & - & \cite{DingCNN} \\
        DNN 2layers  & 841221 & 1680670 & \cite{Params4} \\
        DNN 3layers  & 1235717 & 2469150 & \cite{Params4} \\
        DNN 4layers  & 1366789 & 2731038 & \cite{Params4} \\
        1D-CNN  & 90373 & 6886280 & \cite{Params4} \\
        % \hline
        
        2D CNN  & 1966086 & - & \cite{2dcnn} \\
        IBYOL-IDS  & 1578145 & - & \cite{lightweigh} \\
        CMAE & 130577 & - & \cite{lightweigh} \\
        % \hline
        SS-Deep-ID & 663434 & - & \cite{lightweigh} \\
        E-GraphSAGE  & 94722 & - & \cite{lightweigh} \\
        LSTM-FCNN  & 1137157 & 133300224 & \cite{lightweigh} \\
        RNN & 129157 & - & \cite{lightweigh} \\
        % \hline
        Transformer-CNN-LSTM & 68166 & - & \cite{Transformer} \\
        PyConv & 55877 & - & \cite{PyConv} \\
        \textbf{lw-CNN} & \textbf{49469} & \textbf{729000} & \textbf{Ours} \\
        \hline      
    \end{tabular} 
    }
\end{table}

\vspace{-5pt}

In terms of model size, lw-CNN is 193.24KB, while the Li's CNN is 1.33MB, RNN is 504.52KB \cite{lightweigh}, Transformer-CNN-LSTM is 266.27KB, CatBoost is 1068KB \cite{Transformer} and Aljuaid's CNN is 374.26 KB \cite{aljuaidCNN}. We tested lw-CNN detect time on our machine, average on over 20 tests (batch size 16), resulting in 0.636 ms per sample in NSL-KDD. The lw-CNN has low storage and computation resource requirements. These make lw-CNN suitable for deployment in robots or other automation devices. 
% \vspace{-5pt}

\section{CONCLUSION}
In this paper, we proposed CFedSSL-NID, a federated semi-supervised learning framework for network intrusion detection in Internet of Robotic Things. It utilizes a distributed training paradigm between clients and servers, leveraging data and computational capabilities across distributed robot clients to collaboratively train a robust, accurate and intelligent IoRT DLNIDS. It ensures data privacy preserving for robot clients through federated learning. Semi-supervised based contrastive learning is employed to utilize unlabeled data on robot clients and labeled data on server. The adoption of lw-CNN as both the local and global DLNIDS model facilitates deployment on resource-limited robotic and automation devices. Aforementioned techniques address challenges encountered in practical IoRT. 

Futhermore, CFedSSL-NID enhances performance, generalization, and robustness through various strategies. Randomly weak/strong data augmentation and Dropout can boost model generalization and robustness; Latent contrastive learning can improve performance from unlabeled data, and EMA update can fine-tune the self-supervised parameters with supervised signals.

Deployment of CFedSSL-NID in practical IoRT requires further optimization in complexity (compressing lw-CNN, leveraging accelerators), network comunication (low-latency protocols, bandwidth management), data privacy (encryption, differential privacy) and evaluation in real robots.


%%%%%%%%%%%%%%%%%%%%%%%%%%%%%%%%%%%%%%%%%%%%%%%%%%%%%%%%%%%%%%%%%%%%%%%%%%%%%%%%

%%%%%%%%%%%%%%%%%%%%%%%%%%%%%%%%%%%%%%%%%%%%%%%%%%%%%%%%%%%%%%%%%%%%%%%%%%%%%%%%



% \section*{ACKNOWLEDGMENT}

% The preferred spelling of the word ÒacknowledgmentÓ in America is without an ÒeÓ after the ÒgÓ. Avoid the stilted expression, ÒOne of us (R. B. G.) thanks . . .Ó  Instead, try ÒR. B. G. thanksÓ. Put sponsor acknowledgments in the unnumbered footnote on the first page.

%%%%%%%%%%%%%%%%%%%%%%%%%%%%%%%%%%%%%%%%%%%%%%%%%%%%%%%%%%%%%%%%%%%%%%%%%%%%%%%%

% References are important to the reader; therefore, each citation must be complete and correct. If at all possible, references should be commonly available publications.
\bibliography{ref} % 使用refs.bib文件

\begin{thebibliography}{10}
\providecommand{\url}[1]{#1}
\csname url@rmstyle\endcsname
\providecommand{\newblock}{\relax}
\providecommand{\bibinfo}[2]{#2}
\providecommand\BIBentrySTDinterwordspacing{\spaceskip=0pt\relax}
\providecommand\BIBentryALTinterwordstretchfactor{4}
\providecommand\BIBentryALTinterwordspacing{\spaceskip=\fontdimen2\font plus
\BIBentryALTinterwordstretchfactor\fontdimen3\font minus \fontdimen4\font\relax}
\providecommand\BIBforeignlanguage[2]{{%
\expandafter\ifx\csname l@#1\endcsname\relax
\typeout{** WARNING: IEEEtran.bst: No hyphenation pattern has been}%
\typeout{** loaded for the language `#1'. Using the pattern for}%
\typeout{** the default language instead.}%
\else
\language=\csname l@#1\endcsname
\fi
#2}}

\bibitem{IoRT1}
O.~Vermesan, R.~Bahr, Ottella, and et~al., ``Internet of robotic things intelligent connectivity and platforms,'' \emph{Frontiers in Robotics and AI}, vol.~7, p. 509753, 2020.

\bibitem{Robotresource}
M.~Afrin, J.~Jin, A.~Rahman, Y.-C. Tian, and A.~Kulkarni, ``Multi-objective resource allocation for edge cloud based robotic workflow in smart factory,'' \emph{Future generation computer systems}, vol.~97, pp. 119--130, 2019.

\bibitem{IoRTsensor}
O.~Vermesan, A.~Br{\"o}ring, and et~al., ``Internet of robotic things--converging sensing/actuating, hyperconnectivity, artificial intelligence and iot platforms,'' in \emph{Cognitive hyperconnected digital transformation}.\hskip 1em plus 0.5em minus 0.4em\relax River Publishers, 2022, pp. 97--155.

\bibitem{cui2023novel}
J.~Cui, L.~Zong, J.~Xie, and M.~Tang, ``A novel multi-module integrated intrusion detection system for high-dimensional imbalanced data,'' \emph{Applied Intelligence}, vol.~53, no.~1, pp. 272--288, 2023.

\bibitem{andronie2023big}
M.~Andronie, G.~L{\u{a}}z{\u{a}}roiu, M.~Iatagan, and et~al., ``Big data management algorithms, deep learning-based object detection technologies, and geospatial simulation and sensor fusion tools in the internet of robotic things,'' \emph{ISPRS International Journal of Geo-Information}, vol.~12, no.~2, p.~35, 2023.

\bibitem{mahajan2023automatic}
H.~B. Mahajan, N.~Uke, P.~Pise, M.~Shahade, V.~G. Dixit, S.~Bhavsar, and S.~D. Deshpande, ``Automatic robot manoeuvres detection using computer vision and deep learning techniques: a perspective of internet of robotics things (iort),'' \emph{Multimedia Tools and Applications}, vol.~82, no.~15, pp. 23\,251--23\,276, 2023.

\bibitem{mishra2021internet}
N.~Mishra and S.~Pandya, ``Internet of things applications, security challenges, attacks, intrusion detection, and future visions: A systematic review,'' \emph{IEEE Access}, vol.~9, pp. 59\,353--59\,377, 2021.

\bibitem{IoTSurvey}
M.~Hasan, M.~M. Islam, M.~I.~I. Zarif, and et~al., ``Attack and anomaly detection in iot sensors in iot sites using machine learning approaches,'' \emph{Internet of Things}, vol.~7, p. 100059, 2019.

\bibitem{rahman2020internet}
S.~A. Rahman, H.~Tout, C.~Talhi, and et~al., ``Internet of things intrusion detection: Centralized, on-device, or federated learning?'' \emph{IEEE Network}, vol.~34, no.~6, pp. 310--317, 2020.

\bibitem{MCMAHAN2017}
B.~McMahan, E.~Moore, D.~Ramage, and et~al., ``Communication-efficient learning of deep networks from decentralized data,'' in \emph{Artificial Intelligence and Statistics}, 2017, pp. 1273--1282.

\bibitem{UDA}
Q.~Xie, Z.~Dai, E.~Hovy, T.~Luong, and Q.~Le, ``Unsupervised data augmentation for consistency training,'' vol.~33, pp. 6256--6268, 2020.

\bibitem{Mixmatch}
D.~Berthelot, N.~Carlini, I.~Goodfellow, N.~Papernot, A.~Oliver, and C.~A. Raffel, ``Mixmatch: A holistic approach to semi-supervised learning,'' vol.~32, 2019.

\bibitem{fixmatch}
K.~Sohn, D.~Berthelot, N.~Carlini, Z.~Zhang, H.~Zhang, C.~A. Raffel, E.~D. Cubuk, A.~Kurakin, and C.-L. Li, ``Fixmatch: Simplifying semi-supervised learning with consistency and confidence,'' \emph{Advanced in Neural information processing systems}, vol.~33, pp. 596--608, 2020.

\bibitem{FedCy}
H.~Kassem, D.~Alapatt, P.~Mascagni, A.~Karargyris, and N.~Padoy, ``Federated cycling (fedcy): Semi-supervised federated learning of surgical phases,'' \emph{IEEE transactions on medical imaging}, vol.~42, no.~7, pp. 1920--1931, 2022.

\bibitem{FedMatch}
W.~Jeong, J.~Yoon, E.~Yang, and S.~J. Hwang, ``Federated semi-supervised learning with inter-client consistency \& disjoint learning,'' \emph{arXiv preprint arXiv:2006.12097}, 2020.

\bibitem{FedRGD}
Z.~Zhang, Y.~Yang, Z.~Yao, Y.~Yan, J.~E. Gonzalez, K.~Ramchandran, and M.~W. Mahoney, ``Improving semi-supervised federated learning by reducing the gradient diversity of models,'' in \emph{2021 IEEE International Conference on Big Data (Big Data)}.\hskip 1em plus 0.5em minus 0.4em\relax IEEE, 2021, pp. 1214--1225.

\bibitem{FedCon}
Z.~Long, J.~Wang, Y.~Wang, H.~Xiao, and F.~Ma, ``Fedcon: A contrastive framework for federated semi-supervised learning,'' \emph{arXiv preprint arXiv:2109.04533}, 2021.

\bibitem{SimCLR}
T.~Chen, S.~Kornblith, M.~Norouzi, and G.~Hinton, ``A simple framework for contrastive learning of visual representations,'' in \emph{International conference on machine learning}.\hskip 1em plus 0.5em minus 0.4em\relax PMLR, 2020, pp. 1597--1607.

\bibitem{BYOL}
J.-B. Grill, F.~Strub, F.~Altch{\'e}, C.~Tallec, and et~al., ``Bootstrap your own latent-a new approach to self-supervised learning,'' \emph{Advanced in Neural information processing systems}, vol.~33, pp. 21\,271--21\,284, 2020.

\bibitem{StrongAug}
J.~Liu, H.~Tang, and Y.~Liu, ``Perfect alignment may be poisonous to graph contrastive learning,'' \emph{arXiv preprint arXiv:2310.03977}, 2023.

\bibitem{Dropout}
T.~Gao, X.~Yao, and D.~Chen, ``Simcse: Simple contrastive learning of sentence embeddings,'' \emph{arXiv preprint arXiv:2104.08821}, 2021.

\bibitem{tavallaee2009detailed}
M.~Tavallaee, E.~Bagheri, W.~Lu, and A.~A. Ghorbani, ``A detailed analysis of the kdd cup 99 data set,'' in \emph{2009 IEEE symposium on computational intelligence for security and defense applications}.\hskip 1em plus 0.5em minus 0.4em\relax Ieee, 2009, pp. 1--6.

\bibitem{IoTKDD1}
P.~K. Keserwani, M.~C. Govil, E.~S. Pilli, and P.~Govil, ``A smart anomaly-based intrusion detection system for the internet of things (iot) network using gwo--pso--rf model,'' \emph{Journal of Reliable Intelligent Environments}, vol.~7, no.~1, pp. 3--21, 2021.

\bibitem{IoTKDD2}
J.~Liu, B.~Kantarci, and C.~Adams, ``Machine learning-driven intrusion detection for contiki-ng-based iot networks exposed to nsl-kdd dataset,'' in \emph{Proceedings of the 2nd ACM workshop on wireless security and machine learning}, 2020, pp. 25--30.

\bibitem{IoTKDD3}
M.~Bhavsar, K.~Roy, J.~Kelly, and O.~Olusola, ``Anomaly-based intrusion detection system for iot application,'' \emph{Discover Internet of things}, vol.~3, no.~1, p.~5, 2023.

\bibitem{IoTKDD4}
K.~Albulayhi, Q.~Abu Al-Haija, S.~A. Alsuhibany, A.~A. Jillepalli, M.~Ashrafuzzaman, and F.~T. Sheldon, ``Iot intrusion detection using machine learning with a novel high performing feature selection method,'' \emph{Applied Sciences}, vol.~12, no.~10, p. 5015, 2022.

\bibitem{IoTKDD5}
V.~Kumar, A.~K. Das, and D.~Sinha, ``Uids: a unified intrusion detection system for iot environment,'' \emph{Evolutionary intelligence}, vol.~14, no.~1, pp. 47--59, 2021.

\bibitem{MoCo}
K.~He, H.~Fan, Y.~Wu, S.~Xie, and R.~Girshick, ``Momentum contrast for unsupervised visual representation learning,'' in \emph{Proceedings of the IEEE/CVF conference on computer vision and pattern recognition}, 2020, pp. 9729--9738.

\bibitem{tsne}
L.~Van~der Maaten and G.~Hinton, ``Visualizing data using t-sne.'' \emph{Journal of machine learning research}, vol.~9, no.~11, 2008.

\bibitem{CR}
S.~Laine and T.~Aila, ``Temporal ensembling for semi-supervised learning,'' in \emph{International Conference on Learning Representations}, 2022.

\bibitem{FedProx}
X.~Yuan and P.~Li, ``On convergence of fedprox: Local dissimilarity invariant bounds, non-smoothness and beyond,'' \emph{Advances in Neural Information Processing Systems}, vol.~35, pp. 10\,752--10\,765, 2022.

\bibitem{DingCNN}
Y.~Ding and Y.~Zhai, ``Intrusion detection system for nsl-kdd dataset using convolutional neural networks,'' in \emph{Proceedings of the 2018 2nd International conference on computer science and artificial intelligence}, 2018, pp. 81--85.

\bibitem{Params4}
Z.~Li, C.~Huang, S.~Deng, W.~Qiu, and X.~Gao, ``A soft actor-critic reinforcement learning algorithm for network intrusion detection,'' \emph{Computers \& Security}, vol. 135, p. 103502, 2023.

\bibitem{2dcnn}
G.~Andresini, A.~Appice, L.~De~Rose, and D.~Malerba, ``Gan augmentation to deal with imbalance in imaging-based intrusion detection,'' \emph{Future Generation Computer Systems}, vol. 123, pp. 108--127, 2021.

\bibitem{lightweigh}
Z.~Li and W.~Yao, ``A two stage lightweight approach for intrusion detection in internet of things,'' \emph{Expert Systems with Applications}, p. 124965, 2024.

\bibitem{Transformer}
M.~Tawfik, ``Optimized intrusion detection in iot and fog computing using ensemble learning and advanced feature selection,'' \emph{PloS one}, vol.~19, no.~8, p. e0304082, 2024.

\bibitem{PyConv}
J.~He, X.~Wang, Y.~Song, and Q.~Xiang, ``A multiscale intrusion detection system based on pyramid depthwise separable convolution neural network,'' \emph{Neurocomputing}, vol. 530, pp. 48--59, 2023.

\bibitem{aljuaidCNN}
W.~H. Aljuaid and S.~S. Alshamrani, ``A deep learning approach for intrusion detection systems in cloud computing environments,'' \emph{Applied Sciences}, vol.~14, no.~13, p. 5381, 2024.

\end{thebibliography}

\end{document}